\documentclass[prx,twocolumn,floatfix,nofootinbib,preprintnumbers,superscriptaddress,amsmath,amssymb]{revtex4-1}

\usepackage{xcolor,bm,lineno,multirow,float,times}
\usepackage{physics}
\usepackage{rotating}
\usepackage{verbatim}
\usepackage{graphicx}

\usepackage{natbib}

\def\Ang{\AA$^{-1}$ }

\def\ie{{i.e. }}

\begin{document}
\title{ Quantum Spin Fragmentation in Kagome Ice Ho${_3}$Mg${_2}$Sb${_3}$O${_{14}}$}

\author{Zhiling Dun}
\email{zdun3@gatech.edu}
\affiliation{Department of Physics and Astronomy, University of Tennessee, Knoxville, TN 37996, USA}
\affiliation{School of Physics, Georgia Institute of Technology, Atlanta, GA 30332, USA}

\author{Xiaojian Bai}
\affiliation{School of Physics, Georgia Institute of Technology, Atlanta, GA 30332, USA}
\author{Joseph A.~M. Paddison}

\email{jamp3@cam.ac.uk}
\affiliation{School of Physics, Georgia Institute of Technology, Atlanta, GA 30332, USA}
\affiliation{Churchill College, University of Cambridge, Storey's Way, Cambridge CB3 0DS, United Kingdom}

\author{Nicholas P. Butch}
\affiliation{NIST Center for Neutron Research,  Gaithersburg, MD 20899, USA}

\author{Clarina D. Cruz }
\affiliation{Neutron Scattering Division, Oak Ridge National Laboratory, Oak Ridge, TN 37831, USA}

\author{Matthew B. Stone}
\affiliation{Neutron Scattering Division, Oak Ridge National Laboratory, Oak Ridge, TN 37831, USA}

\author{Tao Hong}
\affiliation{Neutron Scattering Division, Oak Ridge National Laboratory, Oak Ridge, TN 37831, USA}

\author{Martin Mourigal}
\affiliation{School of Physics, Georgia Institute of Technology, Atlanta, GA 30332, USA}

\author{Haidong Zhou}
\email{hzhou10@utk.edu}
\affiliation{Department of Physics and Astronomy, University of Tennessee, Knoxville, TN 37996, USA}
\affiliation{National High Magnetic Field Laboratory, Florida State University, Tallahassee, FL 32310, USA}

\date{\today}

\begin{abstract}
A promising route to realize entangled magnetic states combines geometrical frustration with quantum-tunneling effects. Spin-ice materials are canonical examples of frustration, and Ising spins in a transverse magnetic field are the simplest many-body model of quantum tunneling. Here, we show that the tripod kagome lattice material Ho${_3}$Mg${_2}$Sb${_3}$O${_{14}}$  unites an ice-like magnetic degeneracy with quantum-tunneling terms generated by an intrinsic splitting of the Ho$^{3+}$ ground-state doublet, realizing a frustrated transverse Ising model. Using neutron scattering and thermodynamic experiments, we observe a symmetry-breaking transition at $T^{\ast}\approx0.32$\,K to a remarkable quantum state with three peculiarities: a continuous magnetic excitation spectrum down to $T\approx0.12$\,K; a macroscopic degeneracy of ice-like states; and a fragmentation of the spin into periodic and aperiodic components strongly affected by quantum fluctuations. Our results establish that Ho${_3}$Mg${_2}$Sb${_3}$O${_{14}}$ realizes a spin-fragmented state on the kagome lattice, with intrinsic quantum dynamics generated by a homogeneous transverse field.
\end{abstract} 

\maketitle

\section{Introduction}

Quantum spin liquids are exotic states of magnetic matter in which conventional magnetic order is suppressed by strong quantum fluctuations \cite{Balents_2010}.
Frustrated magnetic materials, which have a large degeneracy of classical magnetic ground states, are often good candidates to observe this behavior. A canonical example is spin ice, in which Ising spins occupy a pyrochlore lattice of corner-sharing tetrahedra \cite{Harris_1997,Bramwell_2001}. Classical ground states obey the ``two in, two out" ice rule for spins on each tetrahedron, and thermal excitations behave as deconfined magnetic charges~\cite{Castelnovo_2008,Fennell_2009}. These fractionalized excitations---also known as magnetic monopoles---interact \textit{via} Coulomb's law and correspond to topological defects of a classical field-theory obtained by coarse-graining spins into a continuous magnetization. In principle, topological \emph{quantum} excitations can be generated by adding quantum-tunneling terms to the classical spin-ice model \cite{Hermele_2004,Savary_2012,Gingras_2014}---e.g., by introducing a local magnetic field transverse to the Ising spins \cite{Moessner_2000,Henry_2014}. A search for pyrochlore materials that realize such quantum spin-ice states has found several promising candidates (see, e.g., \cite{Zhou_2008,Ross_2011,Thompson_2011,Fennell_2012,Sibille_2015,Sibille_2016,Petit_2016,Wen_2017,Lhotel_2017,Sibille_2018,Mauws_2018}. However, important challenges remain, including the determination of the often-complex spin Hamiltonians \cite{Jaubert_2015,Yan_2017,Thompson_2017}, the subtle role that structural disorder may play \cite{Sala_2014,Martin_2017,Mostaed_2017}, and the computational challenges associated with simulations of three-dimensional (3D) quantum magnets \cite{Shannon_2012,Kato_2015}.

\begin{figure}[tbp] 
	\begin{center}
		\includegraphics[width=\columnwidth]{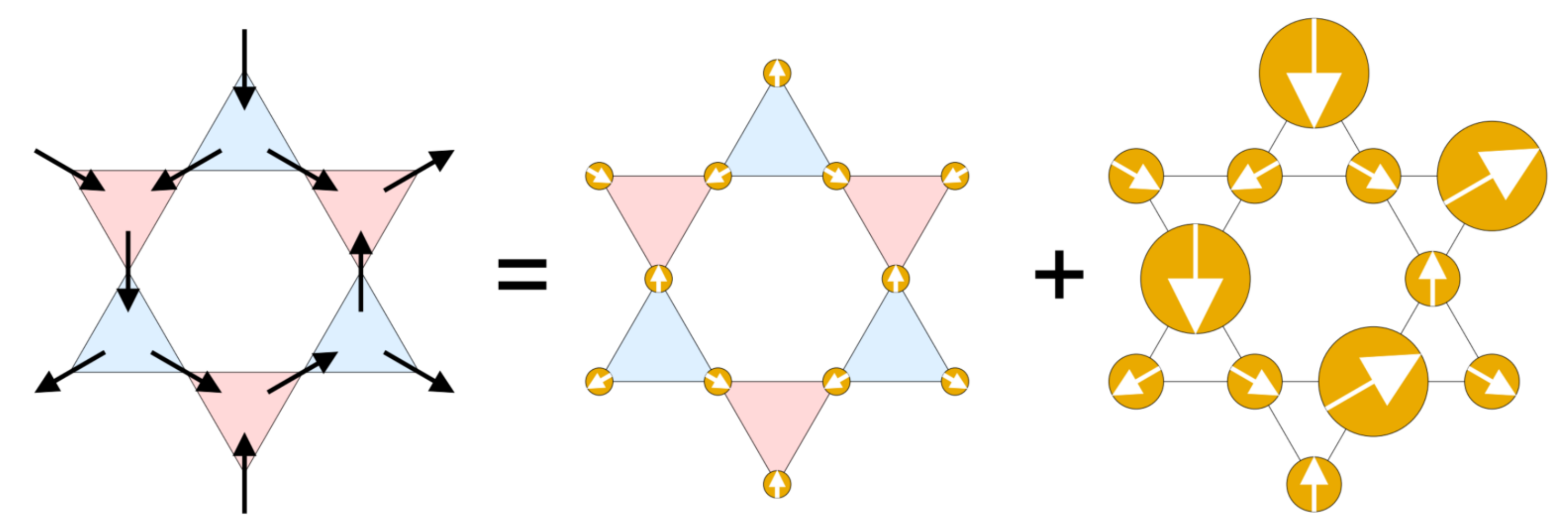}
	\end{center}
	\caption{\label{fig1} Spin fragmentation on a kagome lattice, based on Ref.~\cite{Canals_2016}. Spins are represented by black arrows. Each triangle has two spins pointing ``in" (towards its center) and two pointing ``out" (away from its center). The emergent magnetic charge of a triangle is defined as the number of spins pointing ``in" minus the number pointing ``out". Positive ($+$) and negative ($-$) magnetic charges are represented as red and blue triangles, respectively, and form a staggered arrangement. Such states are macroscopically degenerate because three different spin arrangements are possible for each single charge, two of which are shown in the left image. Spin fragmentation decomposes each unit-length spin into divergence-full and divergence-free channels (center and right images, respectively). The fragmented spins are shown as orange circles with diameter proportional to the length of the fragmented spin. The divergence-full channel corresponds to an ``all-in/all-out" (AIAO) ordering of fragmented spins containing the staggered arrangement of $+$ and $-$ magnetic charges. The divergence-free channel corresponds to a Coulomb phase for which each triangle is charge neutral. Due to the macroscopic degeneracy, the divergence-free part can fluctuate independently of the divergence-full part.}
\end{figure}

A promising alternative route towards quantum analogs of ice is offered by quasi-two-dimensional magnets. In particular, the kagome lattice---a 2D network of corner-sharing triangles---is frustrated for Ising spins coupled by dipolar magnetic interactions \cite{Moller_2009,Chern_2011,Brooks-Bartlett_2014}.  These interactions favor a kagome analog of spin ice with ``one in, two out" and ``two in, one out" spin states on each triangle that carry $\pm1$ emergent magnetic charges~\cite{Wills_2002}. At low temperatures, the effective Coulomb interaction between magnetic charges drives a phase transition to a state with staggered charge ordering \cite{Moller_2009,Chern_2011}. Nevertheless, this state possesses nonzero entropy, because each magnetic charge retains a threefold degeneracy of spin orientations \cite{Moller_2009}; hence, ordering of the magnetic charges does not imply complete ordering of the spins. Fig.~\ref{fig1} shows that such spin structures can actually be decomposed into a divergence-full channel that is spatially ordered, and an divergence-free channel that remains spatially disordered and can fluctuate independently of the divergence-full channel---a process known as spin fragmentation \cite{Brooks-Bartlett_2014,Canals_2016}. Neutron-scattering measurements provide a direct experimental signature of spin fragmentation \emph{via} the coexistence of magnetic Bragg peaks with highly-structured magnetic diffuse scattering \cite{Paddison_2016,Petit_2016,Lefrancois_2017}. In the divergence-free channel, every triangle has zero magnetic charge in a spin-fragmented ground state, and thermal excitations behave as deconfined topological defects, yielding a Coulomb phase analogous to pyrochlore spin ice \cite{Brooks-Bartlett_2014}. Similar to quantum ice states on pyrochlore and square lattices, introducing quantum-tunneling terms in a spin-fragmented phase may likewise generate exotic quantum dynamics on a kagome lattice \cite{Savary_2012,Henry_2014}.

In this work, we show that quantum dynamics exist at the lowest measurable temperatures ($\sim$$0.1$\,K) in the kagome Ising magnet Ho${_3}$Mg${_2}$Sb${_3}$O${_{14}}$ \cite{Dun_2017}. This material is one of an isostructural series of ``tripod kagome" materials derived from the pyrochlore structure by chemical substitution, yielding kagome planes of magnetic rare-earth ions separated by triangular planes of nonmagnetic Mg$^{2+}$ ions [Fig.~\ref{fig2}(a)]. 
Previous neutron-scattering measurements of isostructural Dy${_3}$Mg${_2}$Sb${_3}$O${_{14}}$ revealed a spin-fragmented state at low temperature, in which spin dynamics were unmeasurably slow \cite{Paddison_2016}. In contrast, our measurements of Ho${_3}$Mg${_2}$Sb${_3}$O${_{14}}$ reveal a spin-fragmented state with a broad continnum of low-temperature spin excitations---an experimental signature of quantum dynamics \cite{Han_2012,Paddison_2017}. We explain our experimental results using a model that invokes two symmetry properties: the lower symmetry of the tripod-kagome structure compared to pyrochlore, and the non-Kramers nature of Ho$^{3+}$ ions. Because  time-reversal symmetry does not require non-Kramers ions to have degenerate crystal-field levels, all crystal-field levels are singlets if the point symmetry of the magnetic site is sufficiently low \cite{walter1984treating}. Consequently, the low-energy crystal-field scheme of Ho${_3}$Mg${_2}$Sb${_3}$O${_{14}}$ comprises two singlets, separated by an energy gap of similar magnitude to the magnetic interactions. This two-singlet model maps to an iconic model of \emph{quantum} magnetism---interacting Ising spins in a transverse magnetic field \cite{Wang_1968}---that provides a mechanism for the quantum dynamics we observe. Our results demonstrate that a quantum spin-fragmented state occurs in Ho${_3}$Mg${_2}$Sb${_3}$O${_{14}}$ which has two favorable properties: its quasi-2D structure allows detailed modeling, and its transverse field is intrinsic rather than induced by chemical disorder \cite{Savary_2017,Wen_2017}.

Our paper is structured as follows. In Section~\ref{sec:Methods}, we summarize the experimental and modeling methods that we employ. In Section \ref{sec:TIM}, we introduce the transverse-field Ising model appropriate for Ho${_3}$Mg${_2}$Sb${_3}$O${_{14}}$ at low temperature, motivated by neutron-scattering measurements and point-charge modeling of the crystal-field excitations. In Section~\ref{sec:HC}, we report specific-heat measurements that identify a magnetic phase transition at $T^{\ast}=0.32$\,K. In Section~\ref{sec:HT}, we report inelastic neutron-scattering measurements in the paramagnetic phase above $T^{\ast}$, and show that the paramagnetic spin dynamics obey the form expected for the transverse Ising model. In Section~\ref{sec:LT}, we report low-temperature inelastic neutron-scattering measurements, which show that spin fragmentation occurs below $T^{\ast}$ and a continuum of spin excitations persists in this phase. We show that multi-spin quantum fluctuations beyond mean-field theory are necessary to explain these spin dynamics. Finally, we conclude in Section~\ref{sec:Conclusions} with a discussion of the general implications of our study.

\section{Methods}\label{sec:Methods}
Polycrystalline samples of Ho${_3}$Mg${_2}$Sb${_3}$O${_{14}}$ were synthesized by a solid-state method, following previously-reported procedures \cite{Dun_2016, Dun_2017}. Stoichiometric ratios of Ho$_2$O$_3$ (99.9\%), MgO (99.99\%), and Sb$_2$O$_3$ (99.99\%) fine powder were carefully ground and reacted at a temperature of 1350$^\circ$C in air for 24 hours. This heating step was repeated until the amount of impurity phases as determined by X-ray diffraction was not reduced further. The sample contained a small amount of Ho$_3$SbO$_7$ impurity (2.29(18) wt\%), which orders antiferromagnetically at $T_{\text N}$ = 2.07\,K \cite{Fennell_2001}.

Low-temperature (0.076 $\leq T\leq $ 7\,K) specific-heat measurements were performed in a $^3$He-$^4$He dilution refrigerator using the semi-adiabatic heat pulse technique. The powder samples were cold-sintered with Ag powder. The contribution of the Ag powder was measured separately and subtracted from the data. The lattice contribution to the heat capacity was estimated from measurements of the isostructural nonmagnetic compound La$_3$Zn$_2$Sb$_3$O$_{14}$.

Powder X-ray diffraction measurements were carried out with Cu K$\alpha$ radiation ($\lambda = 1.5418$\,\AA) in transmission mode. Powder neutron-diffraction measurements were carried out using the HB-2A high-resolution powder diffractometer \cite{Garlea_2010} at the High Flux Isotope
Reactor at Oak Ridge National Laboratory, with a neutron wavelength of 1.546\,\AA. Rietveld refinements of the crystal and magnetic structures were carried out using the FULLPROF suite of programs \cite{Rodriguez_1993}. Peak-shapes were modeled by Thompson-Cox-Hastings pseudo-Voigt functions, and backgrounds were fitted using Chebyshev polynomial functions. 

Inelastic neutron-scattering measurements were carried out using the Fine-Resolution Fermi Chopper Spectrometer (SEQUOIA) \cite{Granroth_2010} at the Spallation Neutron Source of Oak Ridge National Laboratory, and the Disk Chopper Spectrometer (DCS) \cite{Copley_2003} at the NIST Center for Neutron Research. For the SEQUOIA experiment, a $\sim$5\,g powder sample of Ho${_3}$Mg${_2}$Sb${_3}$O${_{14}}$ was loaded in an aluminum sample container and cooled to 4\,K with a closed-cycle refrigerator. Data were measured with incident neutron energies of $120$, $60$, and $8$\,meV. The same measurements were repeated for an empty aluminum sample holder and used for background subtractions. For the DCS measurements, the same sample was loaded in a copper can and cooled to millikelvin temperatures using a dilution refrigerator. The measurements were carried out with an incident neutron energy of $3.27$\,meV~at temperatures between $0.12$ and $40$\,K. Measurements of an empty copper sample holder were also made and used for background subtractions. Due to the large specific heat and related relaxation processes below 1\,K, a thermal stabilization time of 6\,h was used; no change in the data obtained was observed after this waiting time. Data reduction was performed using the DAVE program \cite{Azuah_2009}. Data used for fitting were corrected for background scattering using empty-container measurements and/or high-temperature measurements, as specified in the text. These data were also corrected for neutron absorption, placed on an absolute intensity scale by scaling to the nuclear Bragg profile, and the magnetic scattering from the Ho$_{3}$SbO$_{7}$ impurity below its $T_{\text N}$ of $2.07$\,K was subtracted as described in Ref.~\cite{Paddison_2016}.

Point-charge calculations were performed using the software package SIMPRE \cite{SIMPRE} based on a radial effective charge model. The model considers eight effective oxygen charges surrounding a Ho$^{3+}$ ion whose coordinations were defined by the Rietveld refinements to the powder-diffraction data (see Appendix A). The model is then adjusted numerically to match the measured crystal-field spectrum.

For convenience, in the following sections, we use a unit system with $k_\mathrm{B}=1$ and $\hbar=1$, so that all energies are given in units of K.

\section{From crystal structure to transverse field}\label{sec:TIM}

\begin{figure}[tbp] 
	\begin{center}
		\includegraphics[width=\columnwidth]{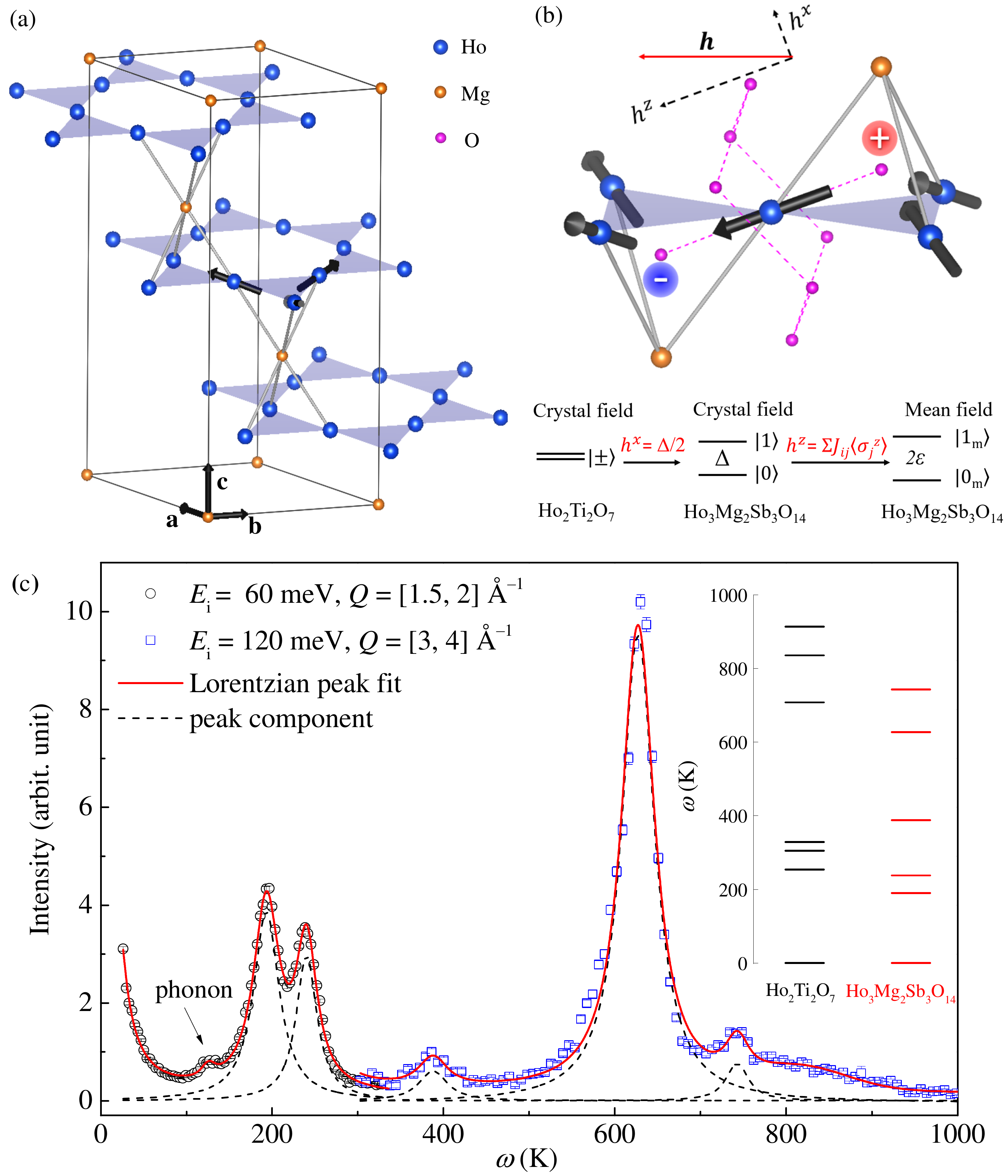}
	\end{center}
	\caption{\label{fig2}{\bf } 
(a) Partial crystal structure of Ho${_3}$Mg${_2}$Sb${_3}$O${_{14}}$, showing alternating Ho$^{3+}$ kagome layers (large blue spheres) and Mg$^{2+}$ triangular layers (small orange spheres). (b)  Local Ho$^{3+}$ environment, showing eight oxygens (small pink spheres), two Mg$^{2+}$ ions, and four nearest-neighbor Ho$^{3+}$ ions around a central Ho$^{3+}$ ion. The three tripod-like Ising axes are enforced by the oxygen in the center of each MgHo$_3$ tetrahedron. Spins are labeled by black arrows and $+$ and $-$ magnetic charges are illustrated by red and blue spheres, respectively. The crystal-field singlets $\ket{0}$ and $\ket{1}$ can be effectively described as a transverse field $h^x$ acting on the non-Kramers doublet ($\ket{\pm}$). The local mean field $h^z$ is a consequence of multi-site magnetic interactions and may mix these singlets into static magnetic states $\ket{0_\mathrm{m}}$ and $\ket{1_\mathrm{m}}$ at low temperature. (c) Crystal-field levels measured by high-energy neutron-scattering experiments. Open black circles and blue squares indicate intensities measured with incident neutron energies of $60$ and $120$\,meV, respectively. Five crystal-field levels are observed at energies of 190(2), 238(2), 388(9), 627(5), and 743(9)\,K, whose peaks are shown as black dashed lines from Lorentzian fits to the data. The overall fit with phonon background scattering is shown as a red line.  Inset: comparison between the observed crystal-field levels at 4\,K of Ho${_3}$Mg${_2}$Sb${_3}$O${_{14}}$ and those of Ho$_2$Ti$_2$O$_7$ (from Ref.~\cite{Ruminy_2016}). Note that whereas crystal field theories predict six doublets with five singlet levels for Ho$_2$Ti$_2$O$_7$, and  17 singlet levels for Ho${_3}$Mg${_2}$Sb${_3}$O${_{14}}$, some excitations(splittings) are not experimentally observed(resolved) due to small neutron scattering cross-sections or limited instrument resolutions.  }
\end{figure} 

The crystal structure of Ho${_3}$Mg${_2}$Sb${_3}$O${_{14}}$ (space group $R\bar{3}m$) is shown in Fig.~\ref{fig2}(a), and contains kagome planes of magnetic Ho$^{3+}$ ions separated by nonmagnetic Mg$^{2+}$ triangular layers \cite{Dun_2017}. The Ho$^{3+}$ site has $C_{2h}$ point symmetry and its local environment contains eight oxygens \cite{Dun_2016, Dun_2017}. The orientations of Ho$^{3+}$ spins are constrained by the crystal field to point along the line connecting Ho$^{3+}$ to its two closest oxygen neighbors, which are situated near the centroids of MgHo$_3$ tetrahedra [Fig.~\ref{fig2}(b)]. 
Rietveld co-refinements to X-ray and neutron powder-diffraction data confirm this crystal structure, and reveal a small amount of Ho$^{3+}$/Mg$^{2+}$ site mixing such that $3.2(2)$\% of Ho$^{3+}$ atomic positions are randomly occupied by Mg$^{2+}$ (see Appendix~A). Hence, the extent of chemical disorder in Ho${_3}$Mg${_2}$Sb${_3}$O${_{14}}$ is less than in its Dy$^{3+}$ analog, where the corresponding value is 6(2)\% \cite{Paddison_2016}. 

High-energy inelastic neutron-scattering measurements on our powder sample of Ho${_3}$Mg${_2}$Sb${_3}$O${_{14}}$ resolve five crystal-field excitations, with energies and relative intensities that resemble those of pyrochlore spin ice Ho$_2$Ti$_2$O$_7$ \cite{Rosenkranz_2000,Ruminy_2016} except for an overall downwards renormalization in energy [Fig.~\ref{fig2}(c)]. This is expected due to the similar local Ho$^{3+}$ environments in the two systems. However, there is a crucial difference between them: the monoclinic $C_{2h}$ symmetry of the Ho$^{3+}$ site in Ho${_3}$Mg${_2}$Sb${_3}$O${_{14}}$ is lower than the trigonal $D_{3d}$ site symmetry in Ho$_2$Ti$_2$O$_7$. 
Consequently, whereas the crystal-field ground-state in Ho$_2$Ti$_2$O$_7$ is a non-Kramers doublet, in Ho${_3}$Mg${_2}$Sb${_3}$O${_{14}}$ all crystal-field levels are necessarily singlets \cite{Dun_2017}. The anticipated two-singlet splitting of doublets in  Ho${_3}$Mg${_2}$Sb${_3}$O${_{14}}$ is too small to observe directly in our high-energy neutron scattering measurements, so we proceed using point-charge calculations,, with effective charges chosen to match the measured crystal-field excitation energies (see Appendix B). This model predicts that the two lowest-energy singlets are separated by an energy gap $\Delta\approx1.7$\,K and are well described by symmetric and antisymmetric combinations of free-ion states,
\begin{eqnarray}
\ket{0} \approx \frac{1}{\sqrt{2}} (\ket{8}+\ket{-8}), \nonumber\\
\ket{1} \approx \frac{1}{\sqrt{2}} (\ket{8}-\ket{-8}), \label{eq:singlets}
\end{eqnarray} 
where $\ket{\pm8}\!\equiv\!\ket{J\!=\!8,J^z\!=\!\pm8}$.
Only these two singlets will be thermally populated at low temperatures, because of their large ($190$\,K) separation from higher crystal-field levels [Fig.~\ref{fig2}(c)]. While both $\ket{0}$ and $\ket{1}$ states are individually nonmagnetic, there is a large matrix element $\alpha=\langle0|\hat{J}_{z}|1\rangle \approx 8$ between them, where $\hat{J}_{z}$ is  the $z$-projection angular momentum operator. This generates a $\sim$$10\,\mu_\mathrm{B}$ dynamic magnetic moment
\begin{equation}
	\boldsymbol{\mu}_{i}= -g\mu_\mathrm{B}\alpha \sigma_{i}^{z}\hat{\mathbf{z}}_{i},
\end{equation}
where $g=\frac{5}{4}$ is the Ho$^{3+}$ Land\'{e} factor, $\sigma^z$ is the $z$ Pauli matrix, and $\hat{\mathbf{z}}_{i}$ is a local Ising axis shown in Fig.~\ref{fig2}(b).

The magnetic Hamiltonian of Ho${_3}$Mg${_2}$Sb${_3}$O${_{14}}$ therefore contains two relevant terms at low temperatures: the two-singlet crystal field $H_{\mathrm{CF}}$, and a pairwise interaction $K_{ij}$ between Ising magnetic moments,

\begin{equation}
\mathcal{H}=\mathcal{H}_{\mathrm{CF}}+\frac{1}{2}\sum_{i,j}K_{ij}\boldsymbol{\mu}_{i}\cdot\boldsymbol{\mu}_{j}.\label{eq:H_moments}
\end{equation}

It is an established result \cite{Wang_1968,Savary_2017} that Eq.~\eqref{eq:H_moments} maps exactly to a transverse-field Ising model (TIM),

\begin{equation}\label{eq:TIM}
\mathcal{H}  =  h^{x}\sum_i \sigma_i^x  +  \frac{1}{2}\sum_{i,j}J_{ij}\sigma_i^z\sigma_j^z,
\end{equation}
where the intrinsic transverse field $h^x=\Delta/2$, and the interaction between Ising spins $J_{ij}=K_{ij}\alpha^{2}\hat{\mathbf{z}}_{i}\cdot\hat{\mathbf{z}}_{j}/g^2\mu_{\mathrm{B}}^2$. Positive values of $J_{ij}$ denote antiferromagnetic interactions between Ising spins, but correspond to ferromagnetic values of $K_{ij}$ because $\hat{\mathbf{z}}_{i}\cdot\hat{\mathbf{z}}_{j}=-0.28$ is negative. 
By analogy with isostructural Dy${_3}$Mg${_2}$Sb${_3}$O${_{14}}$ \cite{Paddison_2016}, we expect that $J_{ij}$ contains a nearest-neighbor exchange term $J$ and the long-range magnetic dipolar interaction $D$,
\begin{equation}
J_{ij} =J_{\mathrm{nn}}\delta_{r_{ij},r_\mathrm{nn}}+Dr_{\mathrm{nn}}^{3}\left[\frac{\hat{\mathbf{z}}_{i}\cdot\hat{\mathbf{z}}_{j}-3(\hat{\mathbf{z}}_{i}\cdot\hat{\mathbf{r}}_{ij})(\hat{\mathbf{z}}_{j}\cdot\hat{\mathbf{r}}_{ij})}{r_{ij}^{3}}\right],  \label{eq:ham}
\end{equation}
where $\delta_{r_{ij},r_\mathrm{nn}}$ is the Kronecker delta function, $r_\mathrm{nn}$ is the distance between nearest-neighbor Ho$^{3+}$ ions, $r_{ij}$ is the distance between ions at positions $\mathbf{r}_i$ and $\mathbf{r}_j$, and $\hat{\mathbf{r}}_{ij}=(\mathbf{r}_{i}-\mathbf{r}_{j})/r_{ij}$. The value of $D =  \mu_{0}\mu^{2} / ({4\pi}{k_{\mathrm{B}}r_{\mathrm{nn}}^{3}}) =  1.29$\,K is  fixed by the crystal structure, and we will obtain an experimental estimate of $J_\mathrm{nn}\sim1$\,K in Section~\ref{sec:HT}. Consequently, the magnetic interactions are comparable in magnitude to the transverse field. In principle, interactions between transverse spin components are also possible, but we expect them to be very weak in Ho${_3}$Mg${_2}$Sb${_3}$O${_{14}}$ because of the strongly Ising character of the magnetic moment [Eq.~\eqref{eq:singlets}], so we do not consider them further.

The TIM defined by Eq.~\eqref{eq:TIM} has been used to model diverse physical phenomena, including ferroelectricity \cite{Brout_1966,Stinchcombe_1973}, superconductivity \cite{Anderson_1958}, quantum information \cite{Suzuki_2012,Dutta_2015}, and quantum phase transitions \cite{Ronnow_2005, Coldea_2010}. The interactions that drive magnetic ordering compete with the transverse field that drives quantum tunneling. In a mean-field picture without frustration, a phase transition occurs to a magnetically-ordered state $\ket{0_{\mathrm{m}}}$ with a static magnetic moment, provided the longitudinal field $h^z$ due to interactions dominates $h^x$. The interplay of frustration and transverse field may generate exotic quantum phases \cite{Moessner_2000, Moessner_2001, Nikolic_2005,Savary_2017}. On the kagome lattice, the TIM with nearest-neighbor antiferromagnetic interactions is predicted to have a quantum-disordered ground state for small $h^x$ \cite{Moessner_2000, Moessner_2001, Nikolic_2005}. 
On the pyrochlore lattice, an external field cannot be applied transverse to all spins simultaneously because different local Ising axes are not coplanar, and an intrinsic transverse field is absent in chemically-ordered pyrochlores. However, transverse fields generated by chemical disorder have been identified as a possible route to pyrochlore QSL states \cite{Savary_2017}, and used to explain the spin dynamics of Pr$_2$Zr$_2$O$_7$  \cite{Wen_2017, Sibille_2018} and Tb$_2$Ti$_2$O$_7$ \cite{Bonville_2011,Petit_2012}. Nevertheless, a potential challenge to modeling such materials is that chemical disorder generates a broad distribution of transverse fields in the sample \cite{Benton_2017}. In this context, a key feature of Ho${_3}$Mg${_2}$Sb${_3}$O${_{14}}$ is that its transverse field is intrinsic to the chemically-ordered structure, and to a first approximation is therefore homogeneous.

\begin{figure} 
	\begin{center}
		\includegraphics[width=\columnwidth]{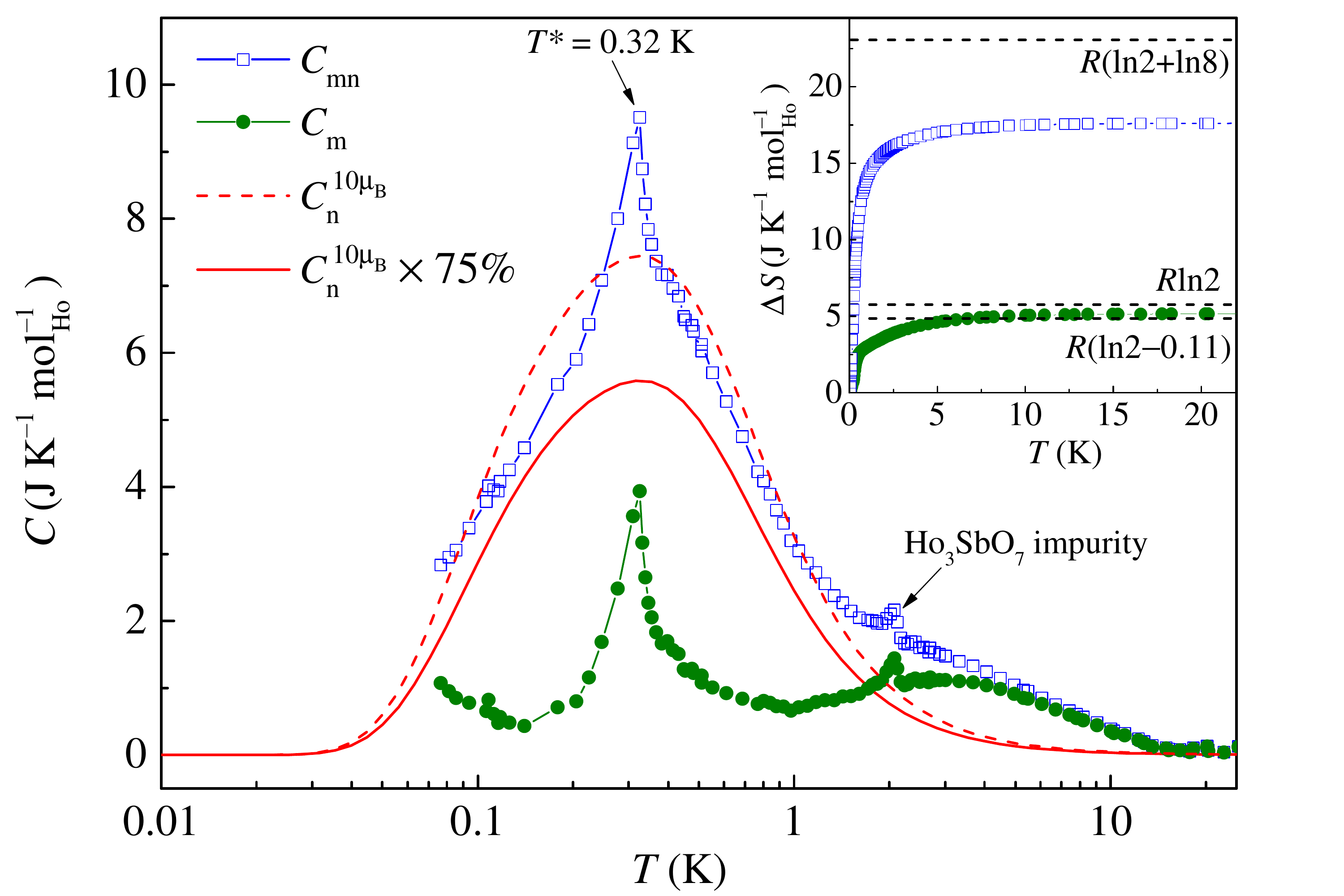}
	\end{center}
	\caption{\label{fig3}  Specific heat of Ho${_3}$Mg${_2}$Sb${_3}$O${_{14}}$, showing the following contributions: (i) sum of the magnetic and nuclear part of the specific heat ($C_{\text{mn}}$, blue squares) obtained by subtracting the lattice contribution from the measured specific heat, (ii) modeled nuclear contribution assuming 100\% static moments of length $10\mu_{\text{B}}$ ($C_{\text{nuc}}^{10\mu_{\text{B}}}$, red dashed line), (iii) modeled nuclear contribution assuming 75\% static moments (red solid line), and (iv) estimated magnetic contribution ($C_{\text{m}}$ = $C_{\text{mn}}-0.75C_{\text{nuc}}^{10\mu_{\text{B}}}$, green circles). Inset: entropy change $\Delta S$ for $C_{\text{mn}}$ (blue squares), and $C_{\text{m}}$ (green circles) from 0.08 to 25\,K. }
\end{figure}

\section{Specific-heat measurements}\label{sec:HC}

We use specific-heat measurements to reveal thermodynamic properties of Ho${_3}$Mg${_2}$Sb${_3}$O${_{14}}$ and identify phase transitions. A sharp peak in the specific heat is observed at $T^\ast = 0.32$\,K, indicating a symmetry-breaking magnetic phase transition [Fig.~\ref{fig3}]. The value of $T^\ast $ is consistent with a broad peak previously reported at $\sim$0.4\,K in the ac susceptibility \cite{Dun_2017}. Although the ac susceptibility peak is frequency dependent \cite{Dun_2017}, the sharpness of the specific-heat peak is inconsistent with a conventional spin-glass transition. The value of $T^\ast$ is also close to the temperature at which isostructural Dy${_3}$Mg${_2}$Sb${_3}$O${_{14}}$ undergoes a spin-fragmentation transition ($\sim$$0.3$\,K in Ref.~\cite{Paddison_2016} and $\sim$$0.37$\,K in Ref.~\cite{Dun_2016}). We will show in Section~\ref{sec:LT} that $T^\ast$ corresponds to the onset of quantum spin fragmentation in Ho${_3}$Mg${_2}$Sb${_3}$O${_{14}}$.

Below $1$\,K, a broad specific heat feature is observed in addition to the sharp peak, consistent with a nuclear contribution. The nuclear specific heat originates from the hyperfine interaction between electronic and nuclear spins, and can be calculated numerically (see Appendix C). In Ho-containing systems with doublet ground states, such as Ho metal \cite{Krusius_1969}, Ho$_2$Ti$_2$O$_7$ \cite{Bramwell_2001PRL}, and LiHoF$_4$ \cite{MENNENGA_1984}, the nuclear specific heat can be modeled by assuming that all electronic spins possess a magnetic moment of 10\,$\mu_\mathrm{B}$/Ho$^{3+}$ that is static on the timescale of spin-lattice nuclear relaxation. In contrast, this assumption strongly overestimates the magnitude of the nuclear specific-heat peak in Ho${_3}$Mg${_2}$Sb${_3}$O${_{14}}$ [Fig.~\ref{fig3}]. This suggests that a fraction of the electronic spins is fluctuating faster than the timescale of spin-lattice nuclear relaxation; i.e., the fraction of static electronic spins $f_{\mathrm{s}}\!<\!1$. We estimate $f_{\mathrm{s}}\approx0.75$ by scaling the calculated nuclear specific heat so that the residual electronic spin contribution to the specific heat is always positive, and the associated electronic spin entropy change between $0.08$ and $25$\,K is close to $R\ln2$, as expected for Ising Ho$^{3+}$ spins [Fig.~\ref{fig3}, inset]. As the large nuclear contribution to the specific heat overlaps with the electronic transition at $T^\ast$, it is not possible to obtain independent estimates of both $f$ and the magnetic entropy change. Nevertheless, the noticeably reduced nuclear contribution compared to other Ho-based compounds provides a first hint of the presence of persistent electronic spin dynamics at low temperatures, similar to the pyrochlore quantum spin-ice candidate Pr$_2$Zr$_2$O$_7$ \cite{Kimura_2013}.

\section{Paramagnetic spin dynamics}\label{sec:HT}

\begin{figure*}[tbp]
	\begin{center}
		\includegraphics[width= 7 in]{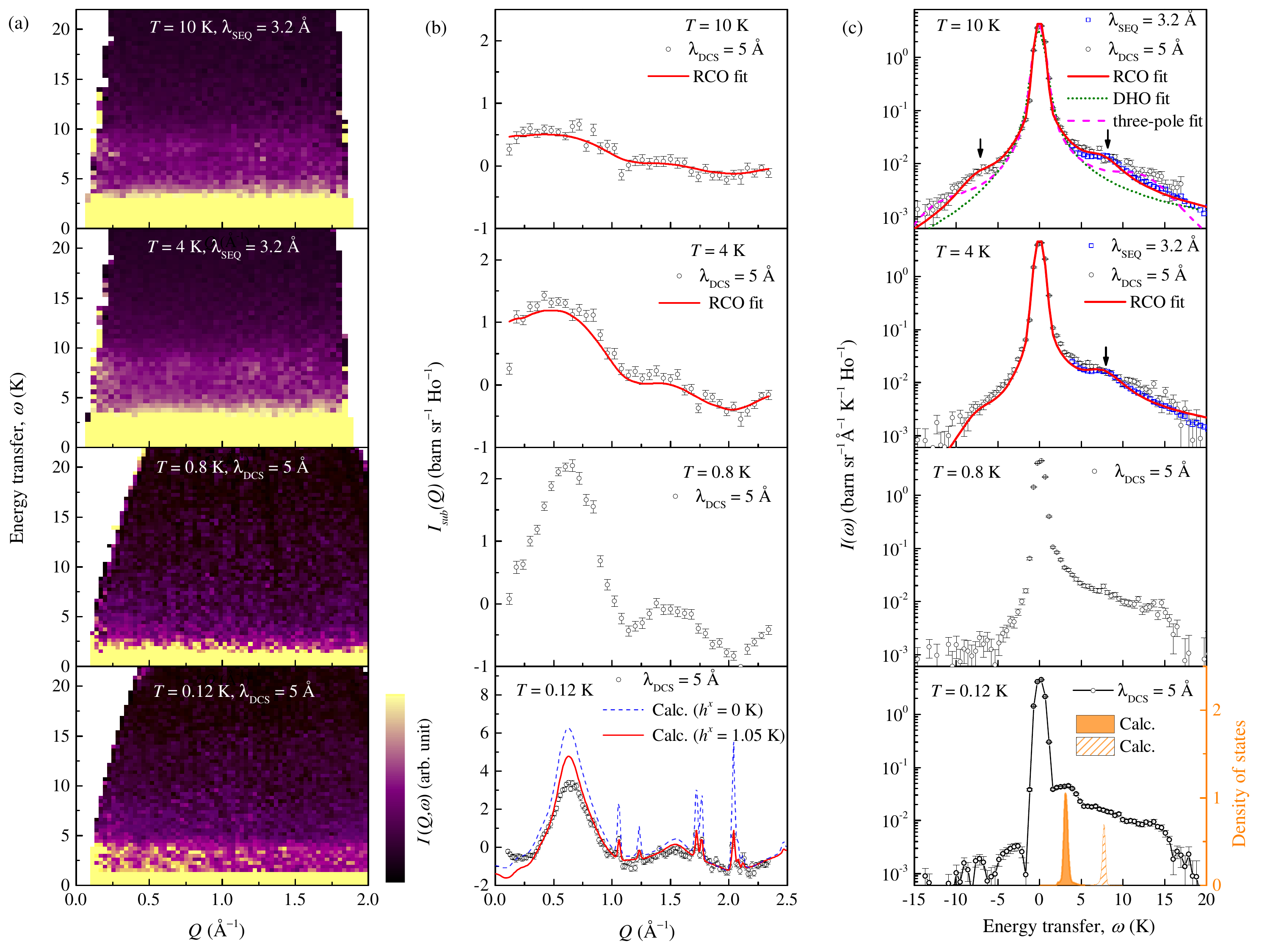}
	\end{center}
	\caption{\label{fig4}  Neutron-scattering data of Ho${_3}$Mg${_2}$Sb${_3}$O${_{14}}$ at temperatures $T=10$\,K, $4$\,K, $0.8$\,K, and $0.12$\,K (top to bottom panels).  (a) Dependence of the inelastic magnetic neutron-scattering intensity $I(Q,\omega)$ on wavevector transfer $Q$ and energy transfer $\omega$ (temperatures as labeled). The upper (lower) two datasets were collected on the SEQUOIA (DCS) instrument with neutron wavelengths of 3.2 (5.0)\,\AA. 
	(b) Energy-integrated ($-25 \leq \omega \leq  25$\,K) magnetic neutron-scattering intensity $I_\mathrm{sub}(Q)$ at four temperatures, showing experimental data collected on the DCS instrument (black circles) and fits to the relaxation-coupled oscillator (RCO) model described in the text (red lines). The correlated magnetic scattering has been isolated by subtracting high-temperature (average of $T=20$ and $40$\,K) data. For $T=0.12$\,K, the calculated pattern for the classical spin-fragmented state is shown as a blue dashed line, and the calculated pattern for the quantum spin-fragmented (QSF) state with $h^{x}=1.05$\,K is shown by the red solid curve (see Section~\ref{sec:LT}).
	(c) Energy dependence of magnetic neutron-scattering intensity $I(\omega)$ at four temperatures, showing data collected on the DCS instrument (black circles) and the SEQUOIA instrument (blue squares), and fits to the RCO model (red lines). Alternative fits to damped harmonic oscillator (DHO, olive dotted line) and three-pole (magenta dashed line)  energy lineshapes are shown in the upper panel, and represent poorer agreement with the experimental data than the RCO model (see Appendix D for details).  Data are integrated over wavevector transfers $0.4 \leq Q \leq1.6 $\,\AA$^{-1}$ and corrected for background scattering using empty-container measurements. For $T = 0.12$\,K, the calculated density of states for single-site excitations of the QSF state is shown by orange peaks.}
\end{figure*}

We use high-resolution inelastic neutron-scattering measurements to further investigate the spin dynamics in Ho${_3}$Mg${_2}$Sb${_3}$O$_{14}$. We consider first the paramagnetic phase above $T^\ast$, and compare our experimental data with calculations for the TIM. At temperatures between $0.8$\,K and $10$\,K, the measured inelastic neutron-scattering intensity shows a broad dependence on energy transfer $\omega$, which is weakly correlated with the dependence on wavevector $Q$ [Fig.~\ref{fig4}(a)]. The $Q$-dependence of the magnetic diffuse scattering was obtained by integrating the inelastic scattering over $-25 \leq \omega \leq  25$\,K. It shows that a broad peak centered at approximately 0.65\,\Ang~develops on cooling the sample, indicating the development of local ice-like correlations \cite{Paddison_2016} [Fig.~\ref{fig4}(b)]. The energy dependence was obtained by integrating the inelastic scattering over $0.4 \leq Q \leq 1.6$\,\AA$^{-1}$. It shows three main magnetic features: an intense resolution-limited central peak; a broad inelastic tail extending to $\omega\approx15$\,K; and a weak inelastic peak at $\omega\approx8$\,K [Fig.~\ref{fig4}(c)]. The presence of significant intensity  in the quasi-elastic channel ($\omega<1$\,K) is qualitatively consistent with the specific-heat results: a majority of spin spectral weight appears static on the timescale of spin-lattice relaxation.

To  interpret the observed spin dynamics, we first discuss the paramagnetic behavior expected for the TIM. If two-site magnetic interactions were absent, the magnetic scattering would contain only a single crystal-field excitation at energy transfer $\Delta=2h^{x}$. However, the presence of interactions strongly modifies this picture. Mean-field theory predicts that interactions split the crystal-field excitation into $n$ dispersive modes, where $n=3$ is the number of magnetic Ho$^{3+}$ ions in the primitive unit cell \cite{Wang_1968}. If the interactions are sufficiently strong compared to the transverse field, they may eventually drive a phase transition to a magnetically-ordered state \emph{via} soft-mode condensation, analogous to the soft phonon modes associated with order-disorder ferroelectric transitions \cite{Brout_1966,Shirane_1971}. Experiments on model two-singlet systems such as LiTbF$_4$ generally support this picture, but show that the dispersive modes are strongly damped \cite{Kotzler_1988,Youngblood_1982,Lloyd_1990}. Subsequent theoretical work proved that this damping is intrinsic to the TIM, and is strongest for $J_{ij} \sim h^x$ \cite{Tommet_1975,Oitmaa_1984,Florencio_1995}. Phenomenologically, the damping can be modeled using ``relaxation-coupled oscillator" (RCO) dynamics, in which dispersive longitudinal modes are coupled to exponentially-relaxing transverse spin components \cite{Kotzler_1988,Lloyd_1990}. Within this framework, the dynamical susceptibility of mode $\mu$ is given by \cite{Kotzler_1988,Shirane_1971}
\begin{equation}\label{eq:RCO}
\chi_{\mu}(\mathbf{Q},\omega) = \frac{4h^{x}\langle\sigma^{x}\rangle} {[\omega_{\mu}(\mathbf{Q})]^{2}-\omega^2-\mathrm{i}\omega\gamma(\omega)},
\end{equation}
where $\omega$ denotes energy transfer, $\langle\sigma^{x}\rangle = \tanh(h^{x} / T)$ is the high-temperature expectation value of the transverse spin, and $[\omega_{\mu}(\mathbf{Q})]^{2}=4h^{x}\{h^{x}-\langle\sigma^{x}\rangle [\lambda_{\mu}(\mathbf{Q})-\lambda]\}$  is the dispersion relation of the $\mu$-th mode derived from an effective-field theory, where $\lambda_{\mu}(\mathbf{Q})-\lambda$ is obtained from the  two-site magnetic interactions (see Appendix D). The damping function $\gamma(\omega)$ obeys the RCO form
\begin{equation}\label{eq:RCO2}
\gamma(\omega) =\Gamma+\frac{\delta^{2}}{\phi-\mathrm{i}\omega},
\end{equation}
where  $\Gamma$ is the damping constant of a dispersive mode, $\phi$ is the energy of a relaxing mode, and $\delta$ is the coupling constant of the two modes \cite{Shirane_1971,Kotzler_1988}.

Inelastic neutron-scattering data directly measure the imaginary part of $\chi(\mathbf{Q},\omega)$, and hence provide a detailed experimental test of validity of the RCO model in Ho${_3}$Mg${_2}$Sb${_3}$O$_{14}$. We find that the RCO model accounts very well for the energy linewidth of the paramagnetic scattering. In contrast, simpler models such as the damped harmonic oscillator disagree with our experimental data [Fig.~\ref{fig4}(c)], because they fail to account for the existence of both a strong central peak and a high-energy tail. This same result was reported for LiTbF$_4$ \cite{Kotzler_1988}, suggesting that the transverse field generates similar paramagnetic dynamics in both materials. A fit to our data indicates a very small relaxing mode energy $\phi$ and in this limit $\chi_{\mu}(\mathbf{Q},\omega)$ reduces to the sum of two terms: a resolution-limited central peak and a damped harmonic-oscillator response with renormalized resonance frequency $[\Omega_{\mu}(\mathbf{Q})]^2 = [\omega_{\mu}(\mathbf{Q})]^2 + \delta^2$. Moreover, our observation of a weak $\omega\approx8$\,K peak can be modeled if a small fraction $f\approx0.1$ of the dispersive modes are underdamped with damping constant $\Gamma_1 < \Omega_{\mu}(\mathbf{Q})$, while the rest are strongly overdamped with a temperature-dependent damping constant $\Gamma_0(T) \gg \Omega_{\mu}(\mathbf{Q})$. 
Considering these effects, our final expression for the scattering function reads (see Appendix D for details), 
\begin{widetext}
\begin{align}
S(\mathbf{Q},\omega)&=\frac{4h^{x}\langle\sigma^{x}\rangle}{N}\sum_{\mu=1}^{N}\left|\mathbf{F}_{\mu}(\mathbf{Q})\right|^{2}\left\{ \frac{T\delta^{2}R(\omega)}{[\omega_{\mu}(\mathbf{Q})]^{2}[\Omega_{\mu}(\mathbf{Q})]^{2}}\right.\nonumber\\&\left.+\frac{\omega}{\pi[1-\exp(-\omega/T)]}\left[\frac{(1-f)\Gamma_{0}}{\left\{ [\Omega_{\mu}(\mathbf{Q})]^{2}-\omega^{2}\right\} ^{2}+(\omega\Gamma_{0})^{2}}+\frac{f\Gamma_{1}}{\left\{ [\Omega_{\mu}(\mathbf{Q})]^{2}-\omega^{2}\right\} ^{2}+(\omega\Gamma_{1})^{2}}\right]\right\}, \label{eq:s_qw_final}
\end{align}
\end{widetext}
where $\mathbf{F}_{\mu}(\mathbf{Q})$ is a structure factor defined in Appendix D, and $R(\omega)$ is the elastic energy resolution function of the experimental data.

We used Eq.~\eqref{eq:s_qw_final} to fit $h^x$, $J_\mathrm{nn}$, $\delta$, $\Gamma_0$, $\Gamma_1(T)$, and $f$ to the momentum, energy and temperature dependence of our paramagnetic neutron-scattering data. The interaction parameter $J_{\mathrm{nn}}$ is mainly constrained by the $Q$ dependence of our data, whereas the other parameters are constrained by its energy dependence. Fits to data are shown in Figs.~\ref{fig4}(b) and \ref{fig4}(c), and fitted parameter values are given in Table~\ref{tab1}. We obtain $J_{\mathrm{nn}}\approx1$\,K, similar to pyrochlore spin ice materials \cite{Hertog_2000}, and $h^x\approx1.5$\,K ($\Delta=3$\,K), which is within a factor of two of the point-charge estimate ($\Delta\approx1.7$\,K). The observed peak at $\sqrt{ (2h^x)^2 +\delta^2}\approx8$\,K is explained by the strong coupling between oscillating modes and exponentially-relaxing transverse spin components. We obtain excellent agreement with experiment at 4\,K and 10\,K, and note that the effective-field theory is not applicable at lower temperatures because of strong short-range spin correlations. This agreement shows that paramagnetic Ho${_3}$Mg${_2}$Sb${_3}$O${_{14}}$ behaves as a canonical TIM, with spin dynamics that resemble model systems such as LiTbF$_4$, and justifies the use of the TIM at lower temperatures. 

\begin{table}[h!]
	\caption{ \label{tab1} Values of the refined parameters from the high-temperature effective-field fits. The Gaussian full width half maximum (FWHM) was fixed to 1.04\,K for the DCS data and 1.73\,K for the SEQUOIA data. Except for $f$, all parameters have units of K.}
	\setlength{\tabcolsep}{5pt}
	\begin{tabular}{ccccccc}
		$J_{\mathrm{nn}}$  & $h^{x}$  & $\delta$  & $\Gamma_{0}^{\textrm{4\,K}}$  & $\Gamma_{0}^{\textrm{10\,K}}$ & $\Gamma_{1}$ & $f$ (\%) \tabularnewline
		\hline 
$0.89(3)$ & $1.48(4)$ & $7.28(6)$  & $79(3)$  & $51(2)$  & $4.7(2)$ & $9.7(5)$ 
	\end{tabular}
\end{table}

\section{Spin fragmentation and quantum dynamics}\label{sec:LT}

We now turn to the low-temperature region for $T < T^*$, where the combination of frustration and quantum tunneling induced by $h^x$ may lead to non-trivial quantum states. Below $T^*$, the most striking feature of our neutron-scattering data is the persistence of continuous magnetic excitations at our lowest measured temperature of $0.12$\,K.  Moreover, these excitations possess additional structure that was absent in the paramagnetic phase. A prominent mode at $\omega \approx3$\,K is now present in $I(Q,\omega)$, as well as a high-energy tail extending to $\omega\approx 15$\,K [Fig.~\ref{fig4}(a) and Fig.~\ref{fig4}(c)]. The presence of low-temperature spin dynamics over a wide energy range strongly contrasts with classical systems such as Ho$_2$Ti$_2$O$_7$ and Dy${_3}$Mg${_2}$Sb${_3}$O${_{14}}$, in which spin dynamics are too slow to observe in neutron-scattering measurements at comparable temperatures \cite{Ehlers_2003,Paddison_2016}.

Our energy-integrated neutron data contain both magnetic Bragg peaks and magnetic diffuse scattering for $T<T^*$ [Fig.~\ref{fig4}(b)]. To explain these data, we first simulate a classical spin-fragmented phase using Monte Carlo simulations of Eq.~\eqref{eq:TIM} with $h^x =0$ \cite{Paddison_2016}. The scattering calculated from this model qualitatively reproduces the relative intensities of the magnetic Bragg peaks and the overall shape of the magnetic diffuse scattering observed experimentally [Fig.~\ref{fig4}(b)]. This result shows that spin fragmentation occurs in Ho${_3}$Mg${_2}$Sb${_3}$O${_{14}}$. However, the observed magnetic Bragg intensities are strongly reduced compared to the classical simulation [Fig.~\ref{fig4}(b)]. The magnetic Bragg intensity is proportional to the square root of the ordered magnetic moment, which is $\mu/3 =3.3\,\mu_\mathrm{B}$ per site for a classical spin-fragmented state in the absence of chemical disorder [Fig.~\ref{fig1}] \cite{Brooks-Bartlett_2014,Paddison_2016}. In contrast, Rietveld refinements to our $0.12$\,K data indicate an ordered magnetic moment of only $1.70(3)$\,$\mu_\mathrm{B}$ per Ho$^{3+}$ (see Appendix A). 
Importantly, Dy${_3}$Mg${_2}$Sb${_3}$O${_{14}}$ has both a larger ordered moment (2.66(6)\,$\mu_\mathrm{B}$ per Dy$^{3+}$ at $0.20$\,K \cite{Paddison_2016}) and a greater degree of site mixing (see Section~\ref{sec:TIM}), which suggests that the observed ordered-moment reduction in Ho${_3}$Mg${_2}$Sb${_3}$O${_{14}}$ cannot be explained by chemical disorder. Instead, the simultaneous enhancement of inelastic scattering and reduction of magnetic Bragg intensity provides experimental evidence for quantum fluctuations.

\begin{figure*}[tbp]
	\begin{center}
		\includegraphics[width= 7 in]{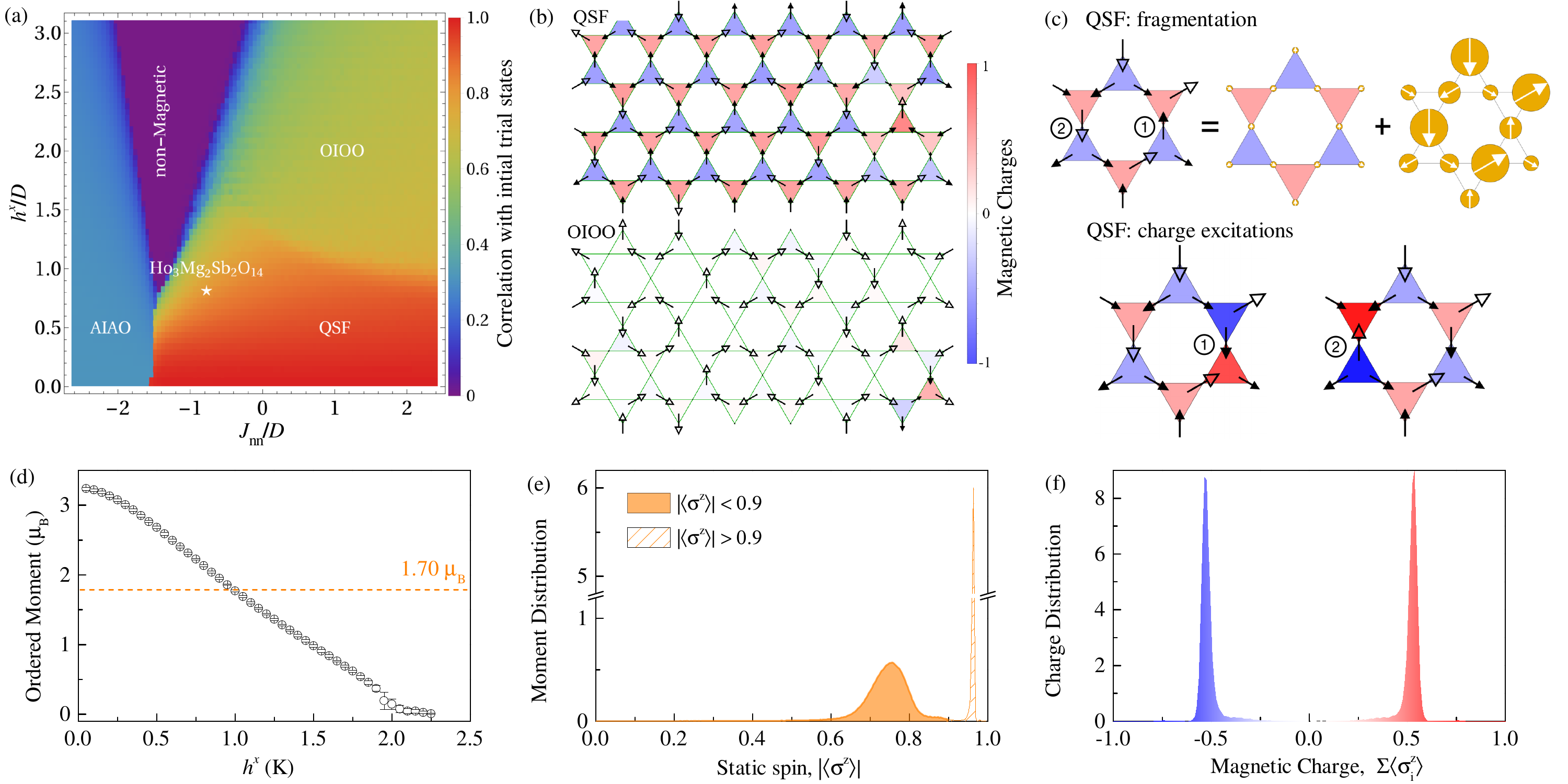}
		\pdfinclusioncopyfonts=1
	\end{center}	\caption{\label{fig5}  Mean-field simulations of our frustrated transverse Ising model. (a) Mean-field phase diagram as a function of $J_{\text{nn}}$/$D$ and $h^{x}$/$D$. The color represents the degree of correlation with the classical spin-fragmented states used as trial states, $\sum_{i}\langle\sigma^{z}_{i}\rangle_{\text{initial}}\langle\sigma^{z}_{i}\rangle_{\text{final}}/N$.  The white star indicates the optimal fitting parameters for Ho${_3}$Mg${_2}$Sb${ _3}$O${_{14}}$ ($h^x$ = 1.05\,K, $J_{\mathrm{nn}}$ = 1.00\,K, $D$ = 1.29\,K). (b) Representative spin configurations in the quantum spin fragmented (QSF) phase and the ``one in, one out" (OIOO) phase. The size of the arrows is scaled with the magnitude of the static spin $\langle\sigma^{z}_{i}\rangle$. Empty arrows denote spins with larger static spins, $|\langle\sigma^{z}_{i}\rangle|>0.9$. Filled arrows denote $|\langle\sigma^{z}_{i}\rangle|<0.9$. (c) Top: illustration of the fragmentation process of the QSF states. Bottom: two types of elementary charge excitations of the QSF state. Single spin flips of the smaller (labeled by number 1) and larger spins (labeled by number 2) are related to the solid and dashed orange areas of the calculated excitation density of states in Fig.~\ref{fig4}(c), respectively.   (d) The ordered moment of QSF states as a function of $h^x$. The orange line corresponds to the observed ordered moment of 1.70\,$\mu_{\mathrm{B}}$ per Ho$^{3}$ at 0.12\,K. (e) Normalized distribution of static spin lengths, $|\langle\sigma^{z}_{i}\rangle|$. (f) Normalized distribution of magnetic charges, defined as the sum of $\langle\sigma_{i}^{z}\rangle$ for each triangle. Both distributions are obtained using the optimal parameters marked by the star in (a).}
\end{figure*}

To gain more insight on the low-temperature state and excitations observed in Ho${_3}$Mg${_2}$Sb${_3}$O${_{14}}$, we use mean-field theory as the simplest starting point to simulate the model of Eq.~\eqref{eq:TIM}.
In the mean-field ground state $\ket{0_\mathrm{m}}=\prod_{i}\ket{0_{\mathrm{m}}}_i$, the expectation value of the static spin at site $i$ is given by \cite{Brout_1966}
\begin{equation}
\langle\sigma^{z}_{i}\rangle =\frac{h^{z}_i}{\sqrt{(h^{x})^{2}+(h^{z}_i)^{2}}}, \label{eq:sigma_z}
\end{equation}
and can vary between $-1$ and $1$. The $z$-component of the mean field arises from two-site magnetic interactions, and is given by
\begin{equation}
\quad h^{z}_i = \sum_{j}J_{ij}\langle\sigma^{z}_{j}\rangle. \label{eq:mf_z}
\end{equation} 
We obtain mean-field equilibrium states by iteratively solving Eqs.~\eqref{eq:sigma_z} and \eqref{eq:mf_z} \cite{Nunez-Regueiro_1992}, using classical spin-fragmented states as initial trial spin configurations (see Appendix~E). The mean-field solution is a product state that includes a full quantum treatment at the level of a single site, but does not capture collective quantum effects such as multi-spin entanglement.

The mean-field phase diagram as a function of $J_\mathrm{nn}$ and $h^{x}$ is shown in Fig.~\ref{fig5}(a). As anticipated, a nonmagnetic singlet ground state with $\langle\sigma^{z}_{i}\rangle=0$ is obtained for large $h^{x}$. For small $h^{x}$ and non-frustrated interactions ($J_\mathrm{nn}/D \ll 0$), a conventionally-ordered ``all-in/all-out" (AIAO) ground state is obtained in which the mean field has the same magnitude on all sites. In contrast, frustrated interactions ($J_\mathrm{nn}/D \gtrsim -1.5$) favor states within the manifold of ``one in, two out" and ``two in, one out" spin configurations, generating a mean field that varies in magnitude from site to site. Consequently, a nonzero transverse field yields mean-field states in which the magnitude of $\langle\sigma^{z}_{i}\rangle$ is spatially modulated ~\cite{Nunez-Regueiro_1992}. This favors two possible frustrated states, depending on the relative strength of $h^{x}$ compared to the magnetic interactions. For larger $h^{x}$, we find a spin-liquid-like phase characterized by a local constraint on every triangle: two of the three static spins form a ``one in, one out" (OIOO) state and trace out closed loops in kagome planes, while the third spin remains entirely dynamic (i.e., the static moment vanishes); hence triangles are magnetic-charge neutral in this state [Fig.~\ref{fig5}(b)]. For smaller $h^{x}$, we obtain a mean-field phase resembling a classical spin fragmented state but dressed with quantum fluctuations; we call this phase \textit{quantum spin fragmented} (QSF) and discuss it in detail below. We note that we could obtain fully periodic state by careful choice of ordered initial configurations (see Appendix E); however, we expect that aperiodic states are entropically favored at finite temperatures due to their macroscopic degeneracy.

In the QSF phase of our mean-field calculations, quantum fluctuations are manifest for each triangle in the form of one ``long" static spin with a large magnitude, and two ``short" static spins with smaller (and possibly different) magnitudes [Figs.~\ref{fig5}(b) and \ref{fig5}(e)]; hence, the shorter spins reflect the presence of persistent dynamics. Remarkably, this state remains spin-fragmented because each triangle has a ``one in, two out" or ``two in, one out" arrangement of static spins that yields a well-defined magnetic charge, and the charges form a staggered arrangement [Figs.~\ref{fig5}(b) and \ref{fig5}(f)]. As for the classical case, ordering of magnetic charges does not imply complete ordering of the static spins; instead, the static spin structures can be decomposed into a divergence-full channel that is spatially ordered and a divergence-free channel that remains spatially disordered. The divergence-full channel has a reduced ordered moment compared to the classical case, because of the intrinsic dynamics associated with the short spins [Fig.~\ref{fig5}(c)]. The divergence-free channel has near-zero magnetic charge on each triangle [Fig.~\ref{fig5}(c)], demonstrating its proximity to a Coulomb phase---an important criterion for a quantum ice-like state \cite{Savary_2012}.

We now compare the behavior of our model in the QSF phase to experimental observations. The ordered moment of $1.70(3)\,\mu_\mathrm{B}$ per Ho$^{3+}$ implies $h^{x}=1.05$\,K, for fixed $J_{\text{nn}}=1.0$\,K [Fig.~\ref{fig5}(d)]. This value is in reasonable agreement with $h^{x}\sim1.5$\,K obtained from fits to our paramagnetic neutron-scattering data (see Section~\ref{sec:HT}). Moreover, simulations of the QSF state with $h^{x}=1.05$\,K show much better qualitative agreement with the measured energy-integrated scattering compared to the classical model with $h^{x}=0$ [Fig.~\ref{fig4}(b)]. Hence, the mean-field QSF state satisfactorily captures the spatial correlations of the experimental system. Turning to the dynamics, the energy gap between the ground state $\ket{0_\text{m}}$ and excited state $\ket{1_\text{m}}$ on a single site is $2\sqrt{(h^{x})^{2}+(h^{z}_i)^{2}}$. The density of magnetic states for the QSF state with $h^{x}=1.05$\,K contains two peaks at $\omega \approx 3.1$\,K and $7.7$\,K [orange areas in Fig. \ref{fig4}(c)], obtained by individual flips of short and long static spins, respectively. These two types of single-spin flips create distinct magnetic-charge excitations: flips of the short spins disrupt the staggered charge arrangement by generating pairs of adjacent triangles with the same charge, whereas flips of the long spins generate a pair of triangles with all-in and all-out spin arrangements [Fig.~\ref{fig5}(c)]. The former excitations qualitatively explain the inelastic mode at $\omega\approx3$\,K in our experimental data, but the latter excitations are not observed experimentally as a distinct mode [Fig.~\ref{fig4}(c)]. Moreover, the mean-field model explains neither the large amount of spectral weight at small energy transfer nor the continuous nature of the excitation spectrum. We therefore conclude that single-site quantum excitations from a mean-field ground state are evidently insufficient to explain the observed spin dynamics.  Collective quantum effects must therefore play a key role in the low-temperature behavior of Ho${_3}$Mg${_2}$Sb${ _3}$O${_{14}}$. We speculate that these correlations may involve quantum-loop dynamics \cite{Hermele_2004,Gingras_2014,Nikolic_2005}, which would allow the system to move between degenerate spin-fragmented ground states and would therefore occur at small energy transfers, as well as deconfinement of charged excitations at higher energies. 

\section{Discussion \& conclusions}\label{sec:Conclusions}
Our experimental study reveal that a quantum spin-fragmented phase occurs at low temperature in Ho${_3}$Mg${_2}$Sb${_3}$O${_{14}}$. 
Our results motivate theoretical calculations to investigate the interplay of spin fragmentation and multi-site quantum tunneling, which will be necessary to explain the continuous low-temperature spin dynamics we observe experimentally.
 Crucially, quantum calculations appear more feasible in Ho${_3}$Mg${_2}$Sb${_3}$O${_{14}}$ than in pyrochlore quantum ices~\cite{Savary_2017}, because of the homogeneous transverse field and the quasi-two-dimensional magnetism of the tripod kagome lattice. Moreover, quantum Monte Carlo modeling of Ho${_3}$Mg${_2}$Sb${_3}$O${_{14}}$ does not suffer from the sign problem, because the interactions involve only the local-$z$ spin components and the transverse field is not frustrated. Experimentally, we anticipate that the application of physical or chemical pressure may drive the system towards a fully fluctuating state, due to its proximity to the phase boundary between quantum-spin fragmented and spin-liquid-like states. 

Our study also highlights the crucial role played by symmetry lowering and long-ranged magnetic interactions. The TIM emerges in Ho${_3}$Mg${_2}$Sb${_3}$O${_{14}}$ because the symmetry of the Ho$^{3+}$ site is lower than that in pyrochlore spin ices. When long-ranged interactions are absent in a TIM, theory predicts that the low temperature states obtained at small transverse fields are continuously connected to the high-field paramagnetic state, and a quantum phase transition to an ice-like state is present only when a small longitudinal magnetic field is applied \cite{Moessner_2000, Moessner_2001, Nikolic_2005}. Long-range dipolar interactions and tripod-like Ising axes in  Ho${_3}$Mg${_2}$Sb${_3}$O${_{14}}$ generate emergent magnetic charges as essential ingredients for realizing kagome ice, through which topological defects are allowed to form and condense without external fields. Consequently, perhaps the most wide-ranging implication of our study is that symmetry lowering and long-ranged interactions need not be a complicating factor in condensed-matter systems, but can actually enable simple models of quantum frustration to be observed experimentally.

\begin{acknowledgements}
We would like to thank Cristian Batista, Owen Benton, Claudio Castelnovo, Laurent Chapon, Radu Coldea, Si\^{a}n Dutton, Michel Gingras, James Hamp, Peter Holdsworth, Ludovic Jaubert, Gunnar M\"{o}ller, Jeffrey Rau, and Han Yan for helpful discussions. The work at the University of Tennessee was supported by the National Science Foundation through award DMR-1350002. H.D.Z acknowledges support from the NHMFL Visiting Scientist Program, which is supported by NSF Cooperative Agreement No. DMR-1157490 and the State of Florida. The work at Georgia Tech is supported by the U.S. Department of Energy, Office of Science, Office of Basic Energy Sciences Neutron Scattering Program under Award Number DE-SC0018660. J.A.M.P. acknowledges financial support from Churchill College, Cambridge (U.K.). The work at NIST was supported in part by the National Science Foundation through award NSF-DMR-0944772. The research at Oak Ridge National Laboratory's Spallation Neutron Source and High Flux Isotope Reactor was sponsored by the U.S. Department of Energy, Office of Basic Energy Sciences, Scientific User Facilities Division.
\end{acknowledgements}

\clearpage
\renewcommand{\thesection}

\section*{Appendix A: Structural and magnetic models}
\begin{figure}[bp]
	\begin{center}
		\includegraphics[width= 3.4 in]{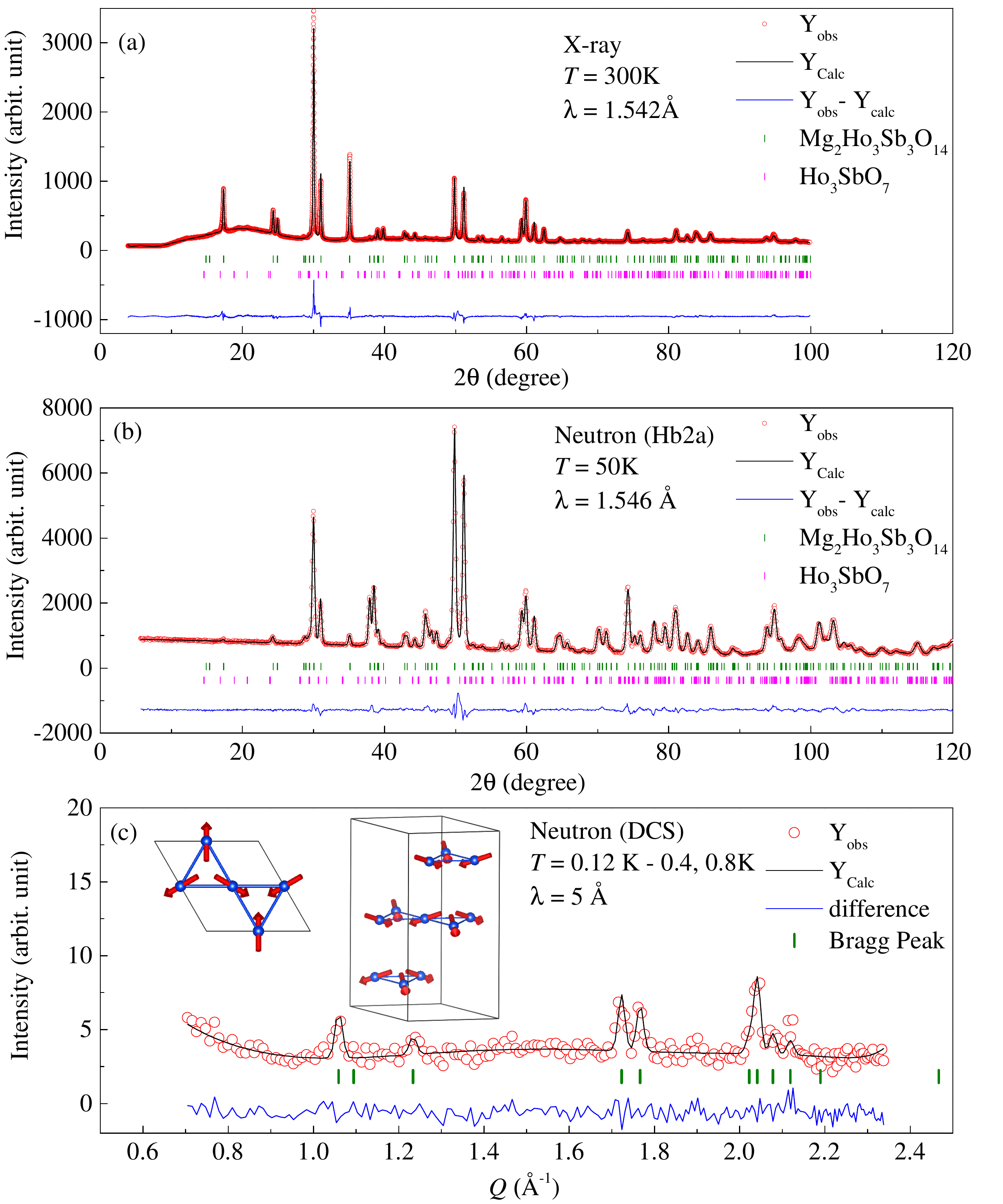}
	\end{center}
	\caption{\label{Sfig:Mag_refine}  Rietveld refinements to diffraction data.  Co-refinements of the crystal structure to neutron and X-ray diffraction data are shown in (a) and (b), respectively. Refinement of the average magnetic structure to low-temperature magnetic diffraction data is shown in (c). In all panels, experimental data are shown as red circles, Rietveld fits as black lines, and difference (data--fit) as blue lines. Inset: illustrations of the all-in/all-out (AIAO) average magnetic structure within a unit cell and a single kagome layer.}
\end{figure}

The structural model of Ho${_3}$Mg${_2}$Sb${_3}$O${_{14}}$ was obtained by Rietveld co-refinements to 50\,K neutron-diffraction data collected using the HB-2A diffractometer [Fig. \ref{Sfig:Mag_refine}(a)] and 300\,K laboratory X-ray diffraction data [Fig.~\ref{Sfig:Mag_refine}(b)]. Refined values of structural parameters, and selected bond lengths and angles are given in Table~\ref{tabS1}. 
The canting angle of the Ising axes with respect to the kagome plane is $22.28(2)^{\circ}$ from the co-refinement.

The average magnetic structure of the QSF state was obtained by Rietveld refinement to energy-integrated neutron-scattering data collected on the DCS spectrometer [Fig.~\ref{Sfig:Mag_refine}(c)]. To isolate the magnetic Bragg scattering below $T^{\ast}$, the difference between data measured at $0.12$\,K, and the average of $0.4$ and $0.8$\,K was used. The average magnetic structure, the AIAO state, belongs to the same irreducible representation as in Dy$_3$Mg$_2$Sb$_3$O$_{14}$, described by $\Gamma_3$ in Kovalev's notation \cite{Kovalev_1993}, consistent with a spin-fragmented state \cite{Paddison_2016}. The refined average moment is $1.70(3)\,\mu_{\text B}$ per Ho$^{3+}$ with a spin canting angle of 24.9$^\circ$ with respect to the kagome plane. 

\begin{table}[tbp]
	\caption{ \label{tabS1} Crystallographic parameters from Rietveld co-refinement to neutron and X-ray diffraction data. 
		Anisotropic atomic displacement parameters were used for Mg1. Fixed parameters are denoted by an asterisk ($\ast$). Selected bond lengths and angles are listed.}
	\begin{center}
		\small
		\renewcommand{\arraystretch}{1}%
		\begin{tabular}{cccccc} 
			\hline
			Atom& Site & $x$ & $y$ & $z$ &  Occ. \\
			\hline
			Mg1 & 3a & 0 & 0 & 0 &  1  \\
			Mg2 & 3b & 0 & 0 & 0.5 &  0.905(7)  \\
			Ho(SD) & 3a & 0 & 0 & 0.5 &  0.095(7)  \\
			Ho    & 9d & 0.5 & 0 & 0.5   &   0.968(2)\\
			Mg(SD)& 9d & 0.5 & 0 & 0.5   & 0.032(2)\\
			Sb    & 9e & 0.5 & 0 & 0     & 1  \\
			O1  & 6c & 0 & 0 & 0.1166(4)  &  1    \\
			O2  & 18h & 0.5214(2) & 0.4786(2) & 0.88960(14)  & 1    \\
			O3  & 18h & 0.4694(2) & 0.5306(2) & 0.35556(13)  &  1    \\
			\hline
			&\multicolumn{5}{c}{Neutron diffraction, $T$ = 50 K}\\
			Lattice para. (\AA)  & \multicolumn{5}{c}{$a$ = $b$ = 7.30195(15), \, $c$ = 17.2569(4) } \\
			$B_\mathrm{an.}(\text {Mg1}) ({\text\AA^{2}})$ & \multicolumn{5}{c}{$B_{\text {11}}$  = $B_{\text {22}}$ = 0.0124(22)} \\
			& \multicolumn{5}{c}{$B_{\text {33}}$ = 0.0002(4),\,  $B_{\text {12}}$ = 0.0062(11) } \\
			$B_\mathrm{iso} ({\text\AA^{2}})$ & \multicolumn{5}{c}{$B(\text{Ho})$ = $B(\text{Sb})$ = 0$\ast$} \\
			 								  & \multicolumn{5}{c}{$B(\text{Mg2})$ = 0.07(13),\, $B(\text{O1})$ = 0.14(8)} \\
											  & \multicolumn{5}{c}{$B(\text{O2})$ = 0.11(5), \,  $B(\text{O3})$ = 0.24(5)  } \\
			Impurity frac. (\%)	& \multicolumn{5}{c} {$f$(Ho$_3$SbO$_7$) = 2.29(18)}\\
			Bond lengths (\AA) & \multicolumn{5}{c}{  Ho--O1 = 2.278(3)} \\
							   & \multicolumn{5}{c}{  Ho--O2 = 2.456(2)} \\
							   & \multicolumn{5}{c}{  Ho--O3 = 2.522(3)} \\
			Bond angles ($^{\circ}$) & \multicolumn{5}{c}{ O1--Ho--O2 = 78.69(10)}\\
									 & \multicolumn{5}{c}{ O1--Ho--O3 =  76.54(17)}\\
			\hline
			&\multicolumn{5}{c}{X-ray diffraction, $T$ = 300 K}\\
			Lattice para. (\AA) & \multicolumn{5}{c}{$a$ = $b$ = 7.30939(13), \, $c$ = 17.2696(3)} \\
			$B_\mathrm{iso} ({\text\AA^{2}})$ & \multicolumn{5}{c}{Overall  $B$ = 1.38(3)} \\
			Impurity frac. (\%) & \multicolumn{5}{c}{$f$(Ho$_3$SbO$_7$) = 0.75(11)} \\
			Bond lengths (\AA) & \multicolumn{5}{c}{  Ho--O1 = 2.280(3)}\\	
			 				   & \multicolumn{5}{c}{  Ho--O2 = 2.458(2)}\\	
			 				   & \multicolumn{5}{c}{  Ho--O3 = 2.524(3)}\\	
			Bond angles ($^{\circ}$)& \multicolumn{5}{c}{ O1--Ho--O2 = 78.68(10)}\\
									& \multicolumn{5}{c}{ O1--Ho--O3 =  76.55(17)}\\
			\hline
		\end{tabular}
	\end{center}
\end{table}

\section*{Appendix B: Point-charge calculations}
Due to the low point symmetry at the Ho$^{3+}$ site, as many as 15 Steven operators are required to describe the crystal-field Hamiltonian of the system \cite{walter1984treating}. As for fitting the crystal-field spectrum using Steven operators, the number of observables from the inelastic neutron scattering measurements are considerable less than the number of fitting variables, making conventional fitting procedures impracticable.  Instead, we calculated the crystal-field levels and wavefunctions from an effective electrostatic model of point charges around a Ho$^{3+}$ ion \cite{SIMPRE}. The coordinates of the oxygen charges was defined by the refined structural model while their effective charges were scanned numerically to match the overall measured inelastic neutron scattering spectrum. By performing calculations for Ho$_2$Ti$_2$O$_7$, we verified that the point-charge model yields a good estimate of the crystal-field levels and their wavefunctions \cite{Rosenkranz_2000}. For Ho${_3}$Mg${_2}$Sb${_3}$O${_{14}}$, our point charge model predicts that the two lowest-energy singlets are separated by 1.7\,K ($h^{x}=0.8$\,K), and their wave-functions in the total angular momentum ($J = 8, J_z = -8,...+8$) basis are given by
\begin{align}
\ket{0} & =  0.688 (\ket{8}+\ket{-8}) - 0.023(\ket{7}-\ket{-7}) \nonumber\\ & + 0.008(\ket{6} + \ket{-6}) - 0.133(\ket{5} - \ket{-5}) \nonumber\\ & + 0.049(\ket{4} + \ket{-4}) + 0.024(\ket{3} - \ket{-3}) \nonumber\\ &  + 0.013(\ket{2} + \ket{-2}) + 0.069(\ket{1} - \ket{-1}) - 0.020 \ket{0}, \nonumber\\
\ket{1} & = 0.691 (\ket{8}-\ket{-8}) - 0.034(\ket{7}+\ket{-7}) \nonumber\\& + 0.008(\ket{6} - \ket{-6}) - 0.120(\ket{5} + \ket{-5}) \nonumber\\& - 0.000(\ket{4}- \ket{-4}) + 0.066(\ket{3} + \ket{-3}) \nonumber \\ & + 0.005(\ket{2} - \ket{-2}) - 0.047(\ket{1} - \ket{-1}) - 0.000 \ket{0} \nonumber.
\end{align} 

\section*{Appendix C: Nuclear specific heat}
\begin{figure}[btp]
	\begin{center}
		\includegraphics[width=\columnwidth]{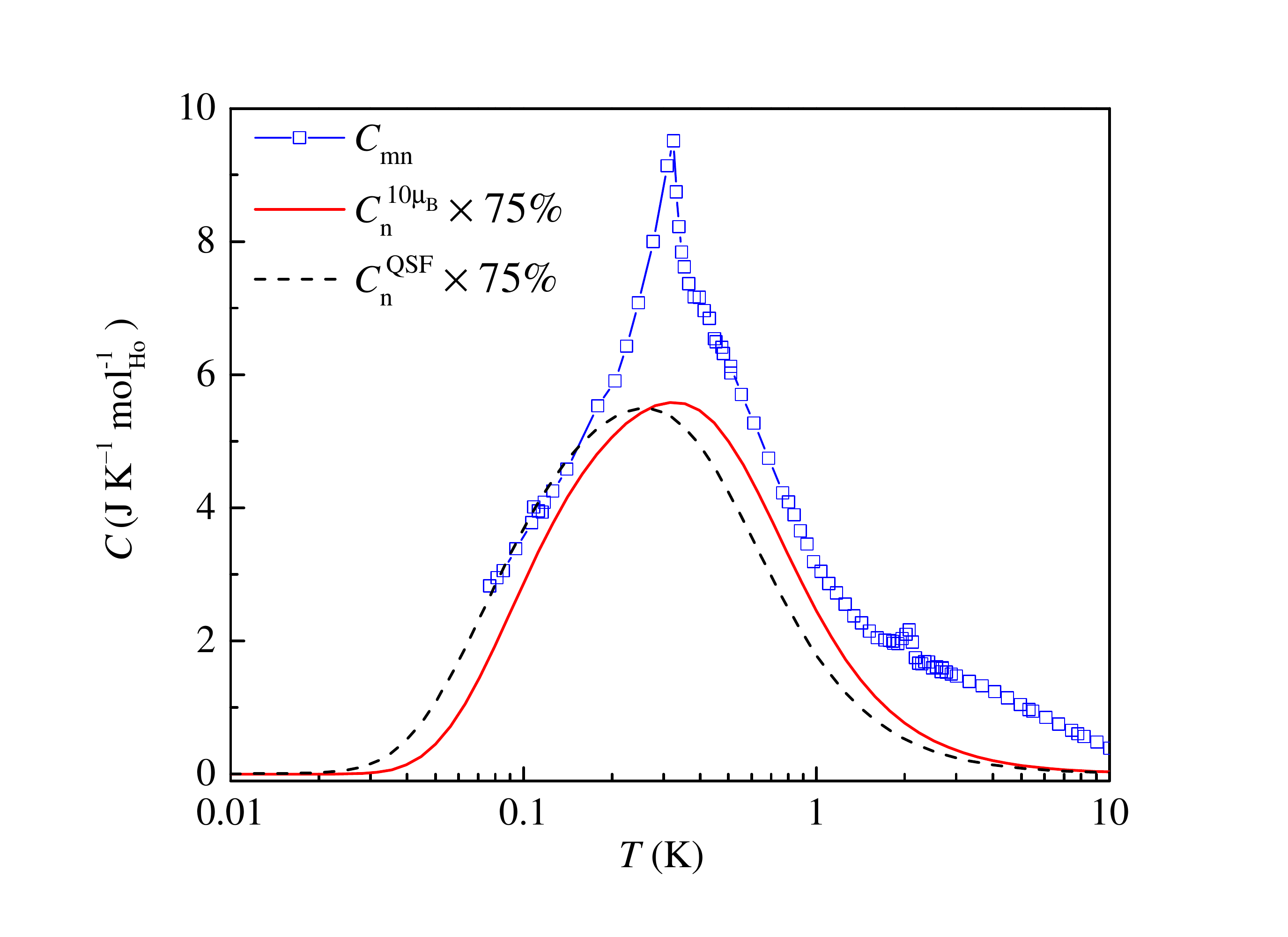}
	\end{center}
	\caption{\label{fig:NucCp}  Calculated nuclear specific heat of Ho$^{3+}$ with full moment of 10\,$\mu_{\text B}$ per Ho$^{3+}$ (red line) and that of the QSF state considering the static moment distribution shown in Fig.~\ref{fig4}(e) (black dashed line). A reduced fraction of static spins $f_{\mathrm{s}}=0.75$ is applied for both cases.}
\end{figure}

The low-temperature nuclear specific heat is given by $f_{\mathrm{s}}C_{\text{n}}$, where $f_{\mathrm{s}}$ is the fraction of static moments, and 
\begin{equation}
C_{\text{n}}=\dfrac{R}{T^2}\dfrac{\sum\limits_{i=-I}\limits^{I}\sum\limits_{j=-I}\limits^{I}\left(W^{2}_{i}-W_{i}W_{j}\right)\exp\left[-(W_{i}+W_{j})/T\right]}{\sum\limits_{i=-I}\limits^{I}\sum\limits_{j=-I}\limits^{I}\exp\left[-(W_{i}+W_{j})/T\right]}\,
\end{equation}
is the nuclear Schottky anomaly.
The energy levels are given by   
\begin{equation}\label{eq:Coupling}
W_{i} = a'i + P\left[i^2-\frac{1}{3}I(I+1)\right]\,, 
\end{equation}
where $I$ is the nuclear spin quantum number, $P$ is the electric quadrupole coupling constant, and $a'=A_{\text{hf}}\dfrac{\mu_{\text{static}}}{g_{J}\mu_{\text{B}}}$ is the hyperfine coupling constant, proportional to the static moment size. Assuming static moments of $10\,\mu_{\text{B}}$ per Ho$^{3+}$, we have $I=7/2$,  $g_{J}=5/4$, $a'=0.319$\,K, and $P=0.0004$\,K \cite{Krusius_1969,kondo1961internal}. The QSF states give rise to a distribution of static moments shown in Fig.~\ref{fig4}(e). We can take this into account when computing the nuclear specific heat by replacing $a'$ with $a'|\langle\sigma_{i}^{z}\rangle|$ and averaging over all the spins, 
\begin{align}
C_{\text{n}}^{\text{QSF}}=\sum_{i}C_{\text{n}}\left(|\langle\sigma_i^{z}\rangle|\right)/N\,.
\end{align}
The result is compatible with the experimental data, assuming a reduced fraction of static spins $f_{\mathrm{s}}=0.75$ [Fig.~\ref{fig:NucCp}]. 

\section*{Appendix D: Paramagnetic effective-field fits}
We used an effective-field approach to calculate the inelastic neutron-scattering
pattern in the paramagnetic phase, based on the Onsager reaction-field
approximation \cite{Santos_1980}. In this approximation, the
inelastic scattering function is given by
\begin{equation}
S(\mathbf{Q},\omega)=\frac{1}{n\pi[1-e^{-\omega/T}]}\sum_{\mu=1}^{N}\left|\mathbf{F}_{\mu}^{\perp}(\mathbf{Q})\right|^{2}\mathrm{Im}\left[\chi_{\mu}(\mathbf{Q},\omega)\right],\label{eq:s_q_w}
\end{equation}
where $n=3$ is the number of Ho$^{3+}$ ions in the primitive unit
cell, $\omega$ is energy transfer in K, and the susceptibility for
each normal mode $\mu$ is given by the RCO form in Eq.~\eqref{eq:RCO2} \cite{Kotzler_1988,Shirane_1971}. The magnetic structure factor is given by
\begin{equation}
\mathbf{F}_{\mu}^{\perp}(\mathbf{Q})=\sum_{i=1}^{N}\mathbf{z}_{i}^{\perp}U_{i\mu}(\mathbf{Q})\exp\left(\mathrm{i}\mathbf{Q}\cdot\mathbf{r}_{i}\right),\label{eq:sf}
\end{equation}
where $\mathbf{Q}$ is the scattering vector, $\mathbf{r}_{i}$ is
the position of magnetic ion $i$ in the primitive cell, and $\mathbf{z}_{i}^{\perp}$
is its local Ising axis projected perpendicular to $\mathbf{Q}$.
The eigenvectors $U_{i\mu}$ and mode energies $\lambda_{\mu}$ are
given at each $\mathbf{Q}$ as the solutions of
\begin{equation}
\lambda_{\mu}(\mathbf{Q})U_{i\mu}(\mathbf{Q})=\sum_{j}J_{ij}(\mathbf{Q})U_{j\mu}(\mathbf{Q}),\label{eq:eigenvalue}
\end{equation}
where the Fourier-transformed interaction $J_{ij}(\mathbf{Q})=\sum_{\mathbf{R}}J_{ij}(\mathbf{R})\exp\left(\mathrm{i}\mathbf{Q}\cdot\mathbf{R}\right)$ includes nearest-neighbor exchange and long-range dipolar contributions,
and $\mathbf{R}$ is the lattice vector connecting atoms $i$ and $j$. The dipolar interaction was calculated using Ewald summation \cite{Enjalran_2004}.
The Onsager
reaction field $\lambda$ is determined by enforcing the total-moment sum rule
\begin{equation}
\frac{1}{nN_{\mathbf{q}}}\sum_{i,\mathbf{q}}\left\langle \sigma_{i}^{z}(\mathbf{q})\sigma_{i}^{z}(-\mathbf{q})\right\rangle =1,\label{eq:onsager_rpa-1}
\end{equation}
which leads to the self-consistency
equation \cite{Santos_1980} 
\begin{equation}
\frac{4h^{x}\langle\sigma^{x}\rangle}{nN_{\mathbf{q}}}\sum_{\mu,\mathbf{q}}\frac{n_{\mu}(\mathbf{q})+\frac{1}{2}}{\omega_{\mu}(\mathbf{q})}=1,\label{eq:onsager_rpa-2}
\end{equation}
where the thermal population factor $n_{\mu}(\mathbf{q})=\left[\exp(\omega_{\mu}(\mathbf{q})/T)-1\right]^{-1}$,
and $\mathbf{q}$ is a wavevector in the first Brillouin zone. We note that Eq.~\eqref{eq:onsager_rpa-2} assumes that the excitations describe delta functions in energy, and we will relax this assumption by considering relaxation effects below. However, we have checked numerically that the total-moment sum rule remains satisfied to within 3\% over the temperature range we consider, so the error introduced by this approximation is small.

The imaginary part of the RCO formula for $\chi_{\mu}(\mathbf{Q},\omega)$
is given by
\begin{equation}
\frac{\mathrm{Im}\left[\chi_{\mu}(\mathbf{Q},\omega)\right]}{4h^{x}\langle\sigma^{x}\rangle}=\frac{\omega(r\phi+\Gamma)}{\left\{ [\omega_{\mu}(\mathbf{Q})]^{2}-\omega^{2}(1-r)\right\} ^{2}+\omega^{2}(r\phi+\Gamma)^{2}},\label{eq:rco_imag_part}
\end{equation}
where $r\equiv\delta^{2}/(\phi^{2}+\omega^{2})$. Preliminary fits
showed that the limit of small $\phi$ (i.e., $\phi\ll\delta^{2}/\Gamma$
and $\phi\ll\left[\omega_{\mu}(\mathbf{Q})\right]^{2}+\delta^{2}$)
was satisfied for our data. In this limit, Eq.~\eqref{eq:rco_imag_part}
reduces to the sum of a damped harmonic oscillator and a Lorentzian central peak \cite{Lloyd_1990},
\begin{align}
\frac{\mathrm{Im}\left[\chi_{\mu}(\mathbf{Q},\omega)\right]}{4h^{x}\langle\sigma^{x}\rangle}=\frac{\omega\Gamma_{0}}{\left\{ [\Omega_{\mu}(\mathbf{Q})]^{2}-\omega^{2}\right\} ^{2}+(\omega\Gamma_{0})^{2}} \nonumber \\+\frac{\delta^{2}}{[\omega_{\mu}(\mathbf{Q})]^{2}[\Omega_{\mu}(\mathbf{Q})]^{2}}\frac{\omega\Gamma_{\mathrm{L}}}{\omega^{2}+\Gamma_{\mathrm{L}}^{2}},\label{eq:RCO_simple-1}
\end{align}
where the pole frequencies of the damped harmonic oscillator are given
by $[\Omega_{\mu}(\mathbf{Q})]^{2}\equiv[\omega_{\mu}(\mathbf{Q})]^{2}+\delta^{2}$,
and the Lorentzian central peak has half-width at half-maximum $\Gamma_{\mathrm{L}}=\phi\omega_{\mu}^{2}(\mathbf{Q})/\Omega_{\mu}^{2}(\mathbf{Q})$
\cite{Lloyd_1990}. We found that $\Gamma_{\mathrm{L}}$ was
smaller than the instrumental energy resolution $R(\omega)$, and therefore replace
the normalised Lorentzian by the instrumental resolution function
and take the limit $\omega\ll T$ for this central peak. To obtain
optimal agreement with experiment, we further assumed that a fraction
$f$ of the system relaxes with damping rate $\Gamma_{1}$ and the
rest with damping rate $\Gamma_{0}$. This yields our final expression
for the scattering function, Eq.~\eqref{eq:s_qw_final}, which was used to obtain the fits shown in Figs.~\ref{fig4}(b) and \ref{fig4}(c). 

The RCO function reduces to simpler models in two limits. First, for $\delta=0$,
Eq.~\eqref{eq:rco_imag_part} reduces to the damped harmonic-oscillator
(two-pole) form,
\begin{equation}
\frac{\mathrm{Im}\left[\chi_{\mu}(\mathbf{Q},\omega)\right]}{4h^{x}\langle\sigma^{x}\rangle}\rightarrow\frac{\omega\Gamma}{[\omega_{\mu}(\mathbf{Q})]^{2}-\omega^{2}+(\omega\Gamma)^{2}}.\label{eq:2-pole}
\end{equation}
Second, for $\Gamma=0$, Eq.~\eqref{eq:rco_imag_part}
reduces to a three-pole form previously proposed for the transverse
Ising model \cite{Tommet_1975},
\begin{align}
&\frac{\mathrm{Im}\left[\chi_{\mu}(\mathbf{Q},\omega)\right]}{4h^{x}\langle\sigma^{x}\rangle}\rightarrow \nonumber \\
&\frac{\omega\phi\delta^{2}}{\phi^{2}\left\{ [\omega_{\mu}(\mathbf{Q})]^{2}-\omega^{2}\right\} ^{2}+\omega^{2}\left\{ [\omega_{\mu}(\mathbf{Q})]^{2}+\delta^{2}-\omega^{2}\right\} ^{2}}.\label{eq:3-pole}
\end{align}
For all fits, the scattering intensities were calculated as
\begin{equation}
I(\omega)=C\left[\dfrac{\mu f(Q)}{\mu_{\text{B}}}\right]^2\int_{Q_{0}}^{Q_{1}}\left\langle S(\mathbf{Q},\omega)\right\rangle _{Q}\mathrm{d}Q,\label{eq:intensity_edep}
\end{equation}
where $Q_{0}=0.4$\,\AA$^{-1}$ and $Q_{1}=1.6$\,\AA$^{-1}$, and
\begin{equation}
I(Q)=C\left[\dfrac{\mu f(Q)}{\mu_{\text{B}}}\right]^2\int_{-\omega^{\prime}}^{\omega^{\prime}}\left\langle S(\mathbf{Q},\omega)\right\rangle _{\omega}\mathrm{d}\omega,\label{eq:intensity_qdep}
\end{equation}
where $\omega^{\prime}=30$\,K, angle brackets here denote numerical spherical
averaging, $f(Q)$ is the Ho$^{3+}$ magnetic form factor \cite{Brown_2004}, $\mu=10\,\mu_\mathrm{B}$ is the total magnetic moment per Ho$^{3+}$, and the constant $C=\left(\gamma_\mathrm{n} r_\mathrm{e}/2\right)^2=0.07265 $\,barn. The integrals were performed numerically.

\section*{Appendix E: Mean-field calculations} 
\begin{figure}[tbp]
	\begin{center}
		\includegraphics[width=\columnwidth]{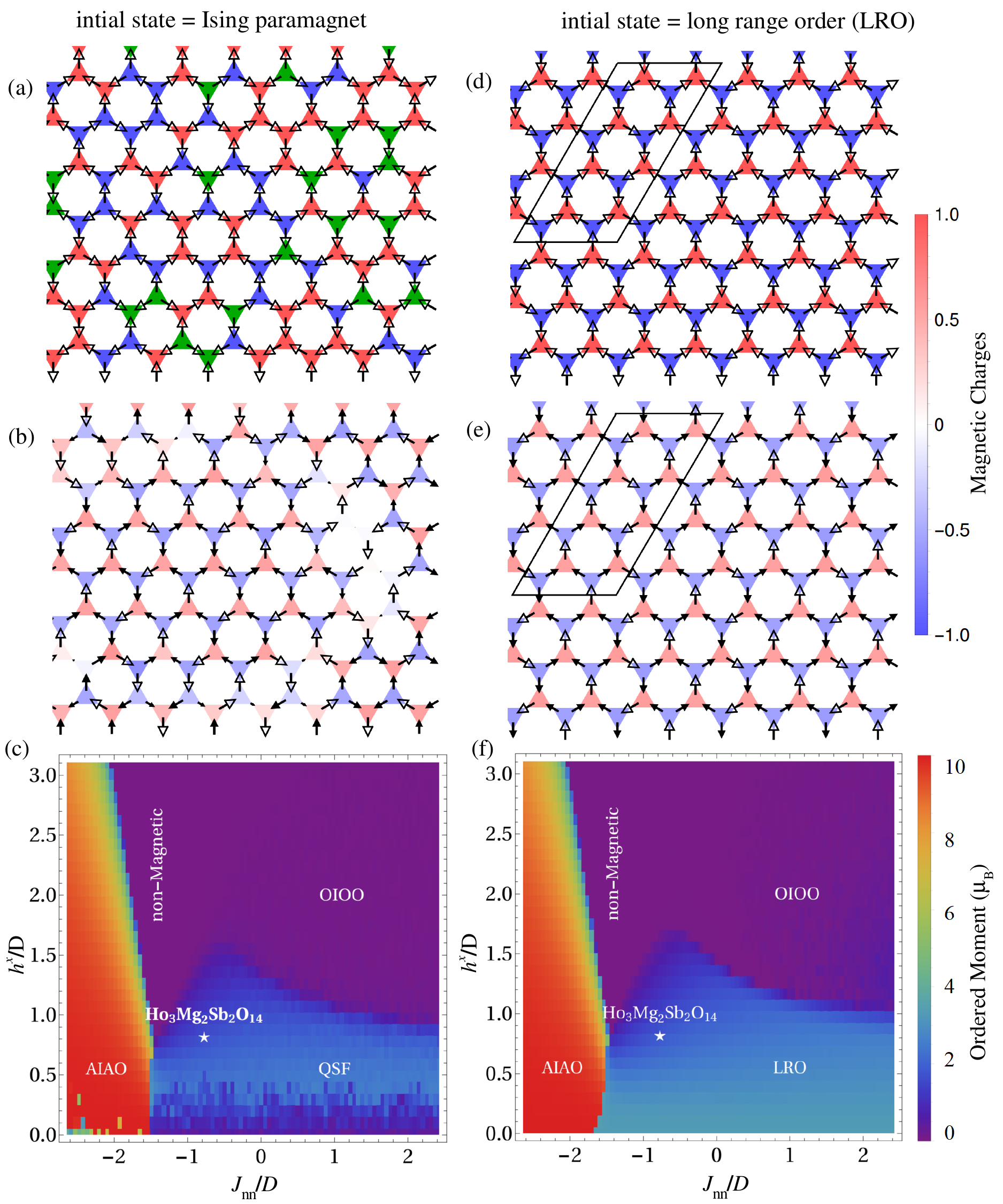}\\
	\end{center}
	\caption{\label{fig:intitial_condition}  Mean-field calculations results using  initial spin configurations other than spin-fragmented states, \ie (a--c) Ising paramagnet and (d--f) long-range ordered states. (a,d) show examples of initial spin configurations in one kagome layer.  (b,e) show the spin configurations of the converged states with parameters for Ho${}_3$Mg${}_2$Sb${}_2$O${}_{14}$ ($h^x = 1.05$\,K, $J_{\mathrm{nn}} = 1.00$\,K, $D= 1.29$\,K).  (c,f) show phase diagrams of ordered magnetic moments for the two initial configurations.  }
\end{figure}

\begin{figure}[tbp]
	\begin{center}
		\includegraphics[width= 3 in]{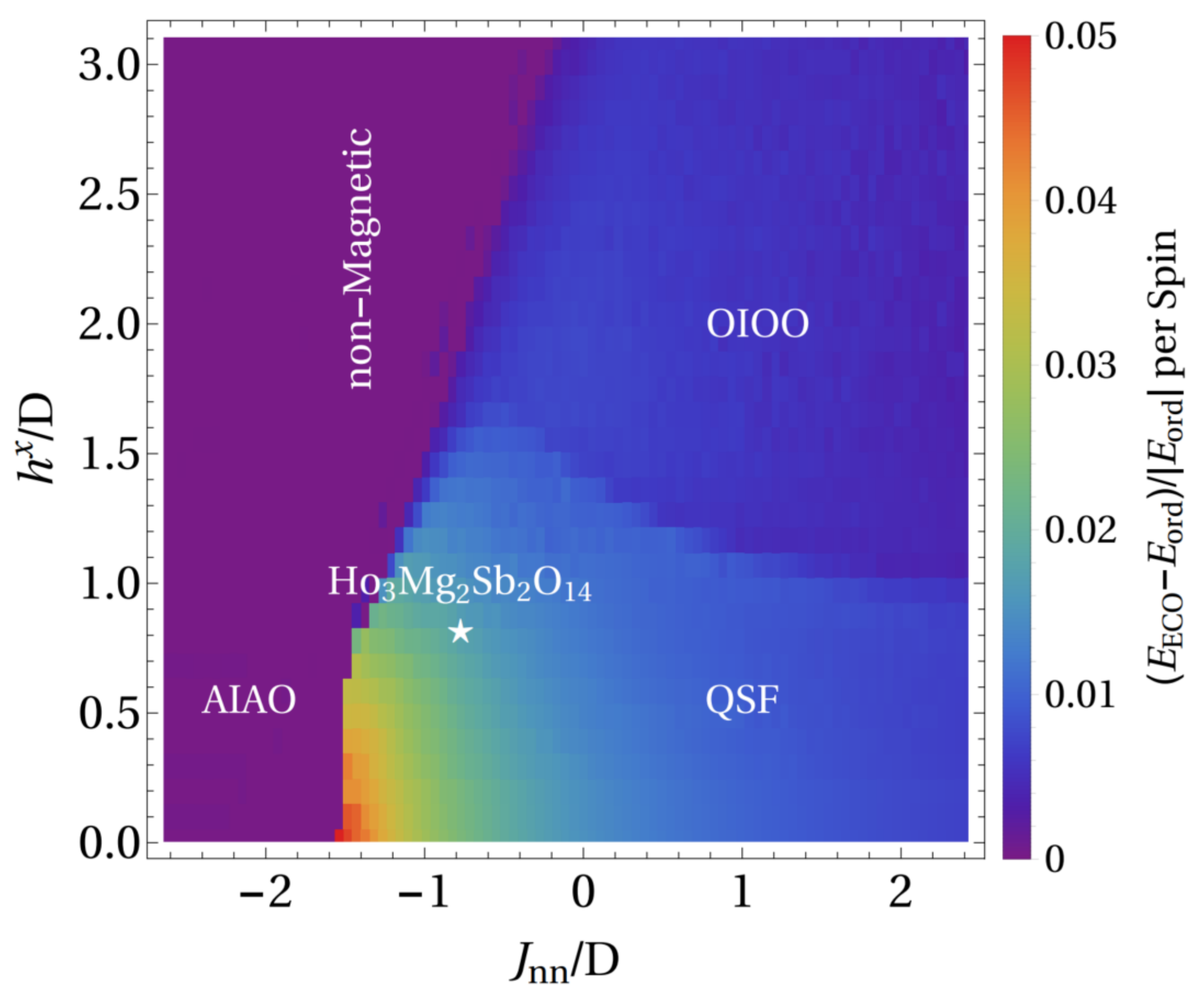}
	\end{center}
	\caption{\label{fig:EngDiff}  Energy difference between the converged mean-field states  using different initial configurations. $E_{\text{ECO}}$ and $E_{\text{ord}}$ denote the energy of the converged mean-field states starting from the classical magnetic charge-ordered configurations and the long-range ordered state, respectively. Different initial configurations converge to the same final state in the non-magnetic phase and the AIAO phase, so there is no energy difference in these regions. For the parameters best describing Ho${}_3$Mg${}_2$Sb${}_2$O${}_{14}$, $h^x$ = 1.05\,K, $J_{\mathrm{nn}}$ = 1.00\,K, $D$ = 1.29\,K, the energy difference is $2.1\%$.}
\end{figure}

We obtain the mean-field Hamiltonian for an arbitrary site $i$ at zero temperature by replacing the operator $\sigma_j^{z}$ in Eq.~\eqref{eq:ham} by its ground state expectation value $\langle\sigma^{z}_{j}\rangle$,
\begin{equation}
{H}^{i}_{\text{MF}}=h^{x}\sigma_{i}^{x}-h^{z}_{i}\sigma_{i}^{z}\,,
\label{eq:MFsite}
\end{equation}
with the local mean-field given by Eq.~\eqref{eq:mf_z} and  ground-state expectation value on that site given by Eq.~\eqref{eq:sigma_z}.
The value of $\sigma_{i}^{z}$ can vary between $-1$ and $+1$. We seek self-consistent solutions to above two equations for a large box of spins using an iterative procedure. Our configurations contain 18 kagome layers and a total of $N=7776$ spins. Starting with an initial distribution of spins, the mean field $h^{z}_i$ is computed at a random site according to Eq.~\eqref{eq:mf_z} and the new spin on that site is updated by Eq.~\eqref{eq:sigma_z}. A total of $N$ such random updates is defined as a sweep. The difference between the old and new configuration is calculated after every sweep and this procedure is repeated until the convergence criterion is met, 
\begin{align}
\sum_{i}|\langle\sigma^{z}_{i}\rangle_{\text{new}}- \langle\sigma^{z}_{i}\rangle_{\text{old}}|/N<10^{-5}\,.
\end{align}  
The interaction matrix $J_{ij}$ consists of nearest-neighbor exchange and long-range dipolar interactions, and is calculated only once at the beginning of the simulation and stored for subsequent computation. The dipolar interaction is treated by Ewald summation with tinfoil boundary conditions at infinity \cite{Leeuw_1980,Melko_2004}, using the formulas for non-cubic unit cells of Ref.~\cite{Aguado_2003}. The effective nearest-neighbor interaction is fixed to be $J_{\text{nn}}=1.00$\,K and the dipolar interaction strength $D=1.29$\,K in all the mean-field calculations. 

We follow Ref.~\cite{Paddison_2016} to calculate the static-moment contribution to the powder-averaged magnetic scattering. The calculated magnetic scattering shown in Fig.~\ref{fig3}(b) is obtained as the sum of static diffuse $I_{\text{diff}}(Q)$, Bragg $I_{\text{Bragg}}(Q)$, and inelastic $I_{\text{in}}(Q)$ contributions, minus the high-temperature paramagnetic $I_{\text{par}}(Q)$ contribution,
\begin{align}
I_\text{sub}(Q)= I_{\text{diff}}(Q)+I_{\text{Bragg}}(Q)+I_{\text{in}}(Q)-I_{\text{par}}(Q),
\end{align}
where the Bragg and diffuse contributions are calculated following Ref.~\cite{Paddison_2016}.
To enforce the total-moment sum rule, the additional inelastic contribution is given by
\begin{align}
I_{\text{inelastic}}(Q) = \dfrac{2}{3}C\left[\dfrac{\mu f(Q)}{\mu_{\mathrm{B}}}\right]^2\dfrac{1}{N}\sum_{i}\left(1-|\langle\sigma_{i}^{z}\rangle|^2\right).
\end{align}

The low-temperature mean-field calculations are mathematically equivalent to solving simultaneous equations in a high-dimensional space, giving rise to many self-consistent solutions that are local minima in energy. Therefore, it is necessary to start from different initial spin configurations (trial states) and calculate the energies of corresponding converged spin configurations (final states). In the absence of transverse fields, the thermodynamics of a classical Ising dipole model on a tripod kagome lattice is described by four temperature regimes: a high-temperature paramagnet that smoothly connects to a short-range order kagome spin ice region, a low-temperature spin-fragmented phase, and an ultra-low temperature long-range spin-ordered phase \cite{Paddison_2016, Chern_2011}. The results shown in Fig.~\ref{fig5} are obtained using the spin-fragmented states as trial states. The calculated results using other physically-meaningful states as initial spin configurations are shown in Fig.~\ref{fig:intitial_condition}.
We first carry out the mean-field iteration from the Ising paramagnetic states where the initial values for $\langle\sigma^z\rangle$ at each site are randomly assigned. By varying  $h^x$ and $J_{\mathrm{nn}}$, an almost identical phase diagram is obtained [Fig.~\ref{fig:intitial_condition}(c)] to that using classical magnetic-charge ordered states as initial states [Fig.~\ref{fig5}(a)]. Within the QSF phase space, some differences are observed due to the formation of charge-ordered domains. 
We also perform the mean-field calculation starting from the ultra-low temperature long-range ordered state. According to classical Monte Carlo simulations, this 3D-ordered state is characterized by a propagation vector $\mathbf{k}=(1/2,1/4,1/2)$, different from the $\sqrt{3}\times\sqrt{3}$ long-range ordered state expected for a single kagome layer \cite{Chern_2011} [Fig.~\ref{fig:intitial_condition}(d)].
The same type of order has also been predicted by Luttinger-Tisza theory \cite{Dun_2016}. 
Using this long-range ordered state as the initial state, a similar phase diagram is obtained [Fig.~\ref{fig:intitial_condition}(f)] with the replacement of the QSF state by a 3D ordered state that also has a modulation in the static spin length [Fig.~\ref{fig:intitial_condition}(c)]. States resulting from the long-range ordered configuration always have lower energy than those converged from the classical magnetic-charge ordered configurations. Therefore, the QSF state is $\textit{not}$ the mean-field ground state of Ho${_3}$Mg${_2}$Sb${_3}$O${_{14}}$.  However, the energy difference between the two states is within $3\%$ [Fig.~\ref{fig:EngDiff}].  Given the small energy difference, the QSF state may be more entropically favorable at finite temperature due to its macroscopic degeneracy.

\bibliographystyle{apsrev4-1}
\bibliography{HoTKL}

\end{document}